\documentclass{aa} 
\usepackage{times,amssymb,amsmath} 
\usepackage{epsfig} 
\usepackage{lscape}
\usepackage[usenames]{color}

\title{The nature of $z \sim 2.3$ Lyman-$\alpha$ emitters
\thanks{Based on observations carried out at the European Southern
Observatory (ESO) under prog. ID No. 084.A-0318(B).}}

\author{K.K. Nilsson\inst{1,2}
        \and G. {\"O}stlin\inst{3}
        \and P. M\o ller\inst{4}
        \and O. M{\"o}ller-Nilsson\inst{2}
        \and C. Tapken\inst{5}
        \and W. Freudling\inst{4}
        \and J.P.U. Fynbo\inst{6}
 }

\institute{ 
 ST-ECF, Karl-Schwarzschild-Stra\ss e 2, 85748, Garching bei M\"unchen, Germany\\
\and
   Max-Planck-Institut f{\"u}r Astronomie, K{\"o}nigstuhl 17,
   69117 Heidelberg, Germany\\	
\and
   The Oskar Klein Centre for Cosmoparticle Physics, Department of Astronomy, Stockholm University, S-106 91 Stockholm, Sweden\\
\and
   European Southern Observatory, Karl-Schwarzschild-Stra\ss e 2, 85748
   Garching bei M\"unchen, Germany\\
\and 
   Astrophysikalisches Institut Potsdam, An der Sternwarte 16,
   14482 Potsdam, Germany \\
\and
   Dark Cosmology Centre, Niels Bohr Institute, University of Copenhagen, 
   Juliane Maries Vej 30, 2100 Copenhagen $\O$, Denmark\\
}
\offprints{knilsson@eso.org}
\date{Received date / Accepted date}
\begin{document}
\titlerunning{The nature of $z\sim2.3$ Ly$\alpha$ emitters}

\abstract{We study the multi-wavelength properties of a set of 171 Ly$\alpha$ emitting candidates at redshift $z = 2.25$ found in the COSMOS field, with the aim of understanding the underlying stellar populations in the galaxies. We especially seek to understand what the dust contents, ages and stellar masses of the galaxies are, and how they relate to similar properties of Ly$\alpha$ emitters at other redshifts. The candidates here are shown to have different properties from those of Ly$\alpha$ emitters found at higher redshift, by fitting the spectral energy distributions (SEDs) using a Monte-Carlo Markov-Chain technique and including nebular emission in the spectra. The stellar masses, and possibly the dust contents, are higher, with stellar masses in the range $\log M_* = 8.5 - 11.0$~M$_{\odot}$ and A$_V = 0.0 - 2.5$~mag. Young population ages are well constrained, but the ages of older populations are typically unconstrained. In 15\% of the galaxies only a single, young population of stars is observed. We show that the Ly$\alpha$ fluxes of the best fit galaxies are correlated with their dust properties, with higher dust extinction in Ly$\alpha$ faint galaxies. Testing for whether results derived from a light-weighted stack of objects correlate to those found when fitting individual objects we see that stellar masses are robust to stacking, but ages and especially dust extinctions are derived incorrectly from stacks. We conclude that the stellar properties of Ly$\alpha$ emitters at $z = 2.25$ are different from those at higher redshift and that they are diverse. Ly$\alpha$ selection appears to be tracing systematically different galaxies at different redshifts.
}

\maketitle

\section{Introduction}
The high redshift Universe hosts a zoo of various types of galaxies, selected with many different methods. Based on numbers of detections, the by far most abundant group of star-forming galaxies with redshifts $z > 3$ are the Lyman-Break Galaxies (LBGs; Steidel et al.~1996, 1999, Madau et al.~1996, Papovich et al.~2001, Giavalisco 2002, Ouchi et al.~2004), found by searching for the Lyman-break in the spectra of galaxies using broad-band imaging. These galaxies often turn out to be moderately star-forming, medium mass galaxies (Shapley et al.~2005; Reddy et al.~2006; Verma et al.~2007). Another large group of high redshift galaxies are found based on their red colours (DRGs, EROs; Franx et al.~2003; Daddi et al.~2004). Unlike the LBGs, and as expected from their colours, these galaxies are often massive and dusty (F{\"o}rster-Schreiber et al.~2004; Papovich et al.~2006). Other types of high redshift galaxies include star-bursting galaxies such as those detected with sub-mm arrays (SMGs; Ivison et al.~1998, 2000, Blain et al.~1999, Chapman et al.~2005) and smaller galaxies detected when gamma-ray bursts occur in them (Bloom et al.~1998, Fynbo et al.~2002, Christensen et al.~2004). One selection method that has become increasingly popular is that of narrow-band imaging of the highly redshifted Ly$\alpha$ line in star-forming galaxies (e.g. M{\o}ller \& Warren~1998, Cowie \& Hu~1998). The largest samples to date include several hundred Ly$\alpha$ emitters between redshifts $z = 2 - 5$ (Gronwall et al.~2007, Venemans et al.~2007, Nilsson et al.~2007, Finkelstein et al.~2007, Ouchi et al.~2008, Grove et al.~2009, Guaita et al.~2010). Slowly, studies of the actual properties of the type of galaxies selected with this method, most notably through SED fitting, are getting under way. 

There is a large spread in the results of previously published work on the topic of Ly$\alpha$ emitter properties and much work is still needed to determine what type of galaxies are selected with the Ly$\alpha$ technique, and where they belong in the high redshift galaxy zoo. To date, eleven publications have presented the SED fitted properties of high redshift Ly$\alpha$ emitters (Gawiser et al.~2006, 2007; Finkelstein et al.~2007, 2009a; Lai et al.~2007, 2008; Nilsson et al.~2007; Pirzkal et al.~2007; Yuma et al.~2010; Ono et al.~2010a, 2010b), and two in the local Universe (Hayes et al.~2007; Finkelstein et al.~2009b). Ly$\alpha$ emitters are selected based on their properties in the UV and have often been found to be extremely faint at longer wavelengths. Since the UV light is dominated by young stars, also contributing to the Ly$\alpha$ emission, little is thus known about the properties of any underlying populations of stars. For galaxies above redshift two, it is thus crucial to have detections of the galaxies in the near- and mid-IR, corresponding to restframe optical and near-IR wavelengths. Few Ly$\alpha$ emitters have been detected in these wavelength regimes, with only a handful to date at $z \sim 3$ (Lai et al.~2008), and at $z = 4 - 6$ (Lai et al.~2007, Pirzkal et al.~2007, Finkelstein et al.~2008, 2009a, Yuma et al.~2010). These studies have shown IR-detected Ly$\alpha$ emitters to be in general more massive, and in some cases more dusty than their non-IR-detected counter-parts. It is not yet clear whether there are indeed two (or multiple) populations of Ly$\alpha$ emitters (e.g. faint-small-dust-free vs.~bright-massive-dusty) or if there is a smooth transition of objects with various masses and properties. 

The surveys to date have concentrated on redshifts $z > 3$. In an effort to get a better handle on the SED of the selected galaxies, we here present the SED study of a sample of Ly$\alpha$ emitters at lower redshift, $z = 2.25$, first presented in Nilsson et al.~(2009a). The advantages of the lower redshift include the possibility of observing a larger range of the restframe spectra with the same instruments, and increased signal-to-noise in the detections due to the smaller luminosity distance. Indications of evolution in the properties of this type of galaxy from redshift $z \sim 3$ to $z \sim 2$ were reported in Nilsson et al.~(2009a); including typically redder colours, smaller equivalent widths and a larger AGN content in the sample. In this paper we investigate the stellar properties of the LAEs further by fitting their SEDs with theoretical models. The question we seek to answer is how these properties have evolved from higher to lower redshift.

This paper is organised as follows: section~\ref{sec:sample} describes the sample of Ly$\alpha$ emitters that are being studied, and includes an analysis of the Spitzer detections of the galaxies. In section~\ref{sec:method} we outline the method used for the SED fitting, and the results of the fitting are reported in section~\ref{sec:results}. A discussion of the redshift evolution of LAE stellar properties is found in section~\ref{sec:discussion}. We end with a general discussion in section~\ref{sec:conclusions} and a conclusion in section~\ref{sec:conclusion}.

\vskip 5mm
Throughout this paper, we assume a cosmology with $H_0=72$
km s$^{-1}$ Mpc$^{-1}$, $\Omega _{\rm m}=0.3$ and
$\Omega _\Lambda=0.7$. Magnitudes are given in the AB system. 

\section{Data}\label{sec:sample}
\subsection{Sample}
The sample studied in this paper is nearly identical to the sample in Nilsson et al.~(2009a). In brief, this sample was found through narrow-band imaging with the ESO2.2m/MPG telescope on La Silla, Chile, using the Wide-Field Imager (WFI; Baade et al. 1999). A central section of the COSMOS field (R.A. $10^h 00^m 27^s$, Dec $02^{\circ} 12' 22$\farcs$7$, J2000) was observed with the filter N396/12, observing Ly$\alpha$ with $z = 2.206 - 2.312$. The data reduction and method for selecting emission line objects is explained in detail in Nilsson et al.~(2009a). In the survey, 187 emission-line objects were found. Of these, 17 were detected in the public GALEX data available and were flagged as potential [OII] emitters, although this sub-sample will also contain Ly$\alpha$ emitters with high escape fraction in the restframe UV. The photometry of these candidates in the optical and near-infrared data available in COSMOS is presented in Nilsson et al.~(2009a). 

In January/February 2010, multi-object spectroscopy with the VIMOS instrument on VLT UT3 (programme ID 084.A-0318(B)) was carried out on 152 candidates. The results of this spectroscopic campaign will be presented in a forthcoming publication, but we will in the analysis here exclude those candidates which are clearly confirmed to not be Ly$\alpha$ emitters in the spectroscopy. Conversely, a number of less likely (named ``unlikely'' in Nilsson et al.~2009a) candidates were put on slits. Those that showed a Ly$\alpha$ detection in the spectroscopy will be added to the analysis here (a total of six). The results of the Nilsson et al.~(2009a) paper do not change significantly from the exclusion/inclusion of the spectroscopic rejections/confirmations. In Table~\ref{tab:sampledef} we present the different sub-samples of the candidates in the various steps. In the final sample that will be used throughout the paper here-after, only candidates that are either not rejected in the spectroscopy, or did not get observed spectroscopically, are included. With ``not rejected'' we take to mean those candidates that were spectroscopically observed and that were left after rejecting galaxies with multiple spectral lines inconsistent with the Ly$\alpha$ redshift, or which did not have significant line emission on a detected continuum. We further add a few more AGN by studying the \emph{Chandra} data in the COSMOS field (see sec.~\ref{sec:chandra}), reducing the number of AGN-free LAE objects. The final sample thus consists of 171 LAEs, of which 141 are AGN- and GALEX-undetected. Note that at least 8 GALEX-detected non-AGN LAEs are confirmed as being true Ly$\alpha$-emitters. These represent a unique sub-sample of high-transmission objects at high redshift (in agreement with M{\o}ller \& Jakobsen~1990), and would be worthy a special follow-up.

\begin{table*}[t]
\begin{center}
\caption{Definitions of all sub-sets of LAE candidates. }
\begin{tabular}{@{}lccccccccc}
\hline
\hline
Sample  & Total & LAE best & GALEX only & AGN only & GALEX + AGN   \\
\hline
Original & 187  & 156 & 12 & 14  & 5   \\
Original (C-list) & 61  & --- & --- & ---  & ---   \\
\hline
Spec. obs. & 118 & 96 & 11 & 8 & 3 \\
Spec. obs. (C-list) & 34 & --- & --- & --- & --- \\
Included f. spec. & 96 & 83 & 8 & 4 & 1 \\
Included f. spec. (C-list) & 6 & 6 & 0 & 0 & 0 \\
\hline
Total after spec. & 171 & 149 & 9 & 10 & 3 \\ 
\hline
\hline
\textbf{Final (after new AGN)} & \textbf{171} & \textbf{141} & \textbf{7} & \textbf{18} & \textbf{5} \\
\hline
\hline
Fitted & 58 & 54 & 4 & --- & --- \\
Good $\chi^2_r$ & 40 & 38 & 2 & --- & --- \\
\hline
\label{tab:sampledef}
\end{tabular}
\end{center}
\begin{list}{}{}
\item[] Each line defines a further sub-sample of the original Nilsson et al.~(2009a) sample (first lines; divided into the originally published sample and the C-list, or ``unlikely'' candidates from that paper), divided into the total sample, the sample without GALEX and AGN detections, and the GALEX only, AGN only and combined GALEX and AGN detected LAEs. The ``unlikely'' candidates were not analysed for AGN and GALEX detections prior to spectroscopic confirmation. The second to fifth rows show the number of candidates put on slits, and those that were not excluded. The row titled ``Total after spec.'' indicates the resulting sample after the spectroscopy, summing those objects that are accepted by the spectroscopy with those without spectroscopic data, and the following row indicates the result of the new AGN detections (sec.~\ref{sec:chandra}). The final rows show how many of each sub-sample were fitted for and which had good $\chi^2_r$ in the fit (sec.~\ref{sec:indresults}). 
\end{list}
\end{table*}

\subsection{Updated photometry}
Since the publication of the original sample, several new images have become available from the COSMOS consortium, including J images from UKIRT, K$_s$ images from CFHT and X-ray data from the \emph{Chandra} observatory (see sec.~\ref{sec:chandra}). The depths of the infrared images are 23.7 magnitudes ($5\sigma$) in both bands (Ilbert et al. 2009). We prepared and extracted fluxes of all the objects in the catalogue in a similar way as described in Nilsson et al.~(2009a). The number of detections was 43 and 86 of the 171 candidates respectively in the J and K$_s$ bands. One of the spectroscopically confirmed, less likely candidates was detected in the K$_s$ band. These results were incorporated into our main photometric catalogue. In the case that an object was already detected in the previously available, but shallower, KPNO K$_s$ band, this magnitude was replaced with the CFHT result for consistency. 

For all optical and near-IR bands used in the fitting (Bj, Vj, $i$, $z$, J, and Ks), the curves of growth of 25 stars in the field were measured and aperture corrections from $3''$ diameter apertures to ``infinity'' calculated. These were applied to all magnitudes before the SED fitting. We also subtracted the observed Ly$\alpha$ flux (as calculated from the narrow-band observations, Mag$_{3960}$) from the Bj band, as this line is exceptionally difficult to fit.

\subsection{Chandra detections}\label{sec:chandra}
Elvis et al.~(2009) presented the available \emph{Chandra} X-ray images, with flux limits of 1.9, 7.3 and 5.7$ \times 10^{-16}$~erg~s$^{-1}$~cm$^{-2}$ in the soft ($0.5-2$~keV), hard ($2-10$~keV) and total ($0.5-10$~keV) bands respectively. To search for detections, aperture photometry with $5''$ diameter was made on the positions of the candidates in the $0.5-10$~keV band. For detections at the $3\sigma$ level, the photometry was repeated in the soft and hard bands. This revealed 19 detections, of which five are only detected in the total band, five/two are detected in the total and soft/hard bands. Seven candidates are detected in all bands. Based on the 1761 sources found in the whole 0.5~deg$^2$ COSMOS survey by Elvis et al.~(2009), the random chance of an object in one aperture is 0.0053, or 0.9 in the whole sample of 171 objects.  All LAEs detected with Chandra (or previously with XMM or VLA, see Nilsson et al.~2009a) are considered AGN due to their bright X-ray (or radio) fluxes. The faintest limit in the X-ray data is that of the soft band from \emph{Chandra}, and this limit corresponds to a SFR of $\sim 1500$~M$_{\odot}$~yr$^{-1}$ according to the conversion factor of Ranalli et al.~(2003). Similarly, the faintest radio detection corresponds to a SFR of $\sim 950$~M$_{\odot}$~yr$^{-1}$ according to the conversion of Bell~(2003). These limits are much higher than those derived from the Ly$\alpha$ line in these galaxies, with the conclusion that all X-ray/radio detections must be AGN-driven. Their R band-to-X-ray fluxes are in the range of what is commonly found for AGN (Schmidt et al. 1998, Hornschemeier et al. 2001). Hence we now have a total of 23 LAE AGN, 10 of which are only detected with \emph{Chandra}, three are only detected in radio emission with VLA, one is only detected with XMM, and 9 that are detected by both \emph{Chandra} and VLA/XMM. The new AGN detections are LAE\_COSMOS\# 14, 20, 52, 67, 76, 120, 130, 134, 151 and 175. 

Thus the confirmed AGN contamination in the non-GALEX-detected sample increases from 9\% (before \emph{Chandra} detections but after spectroscopic rejections) to at least 11\%, i.e. 18/159 (see also Table~\ref{tab:sampledef}), a much larger percentage than in high redshift samples (Wang et al. 2004, Gawiser et al. 2007). Lower redshift samples of Ly$\alpha$ emitters at $z = 0.3$ and $z=2.2$ have also been shown to contain large AGN fractions, of e.g. $\sim 43$\% at $z=0.3$ (Finkelstein et al.~2009c) and $\sim 75$\% at $z=2.2$ (Bongiovanni et al.~2010) indicating a clear redshift evolution, although the result of Bongiovanni et al.~(2010) is based on detecting AGN with a method different from other groups, which may influence the reliability of this fraction in comparison with other results. The increase in AGN fraction was discussed already in Nilsson et al.~(2009a), and we here repeat their test of studying only a sub-sample of the LAEs that are bright enough to have X-ray detections if they are AGN. Using the same quasar composite spectrum as in Nilsson et al.~(2009a), the limiting magnitude in Ly$\alpha$ is 24.79 based on the new X-ray detection limit of the \emph{Chandra} observations. Of the final sample in this publication, 53 LAEs fulfill this criterion. Based on these LAEs only, the AGN fractions are $22\pm7$\% and $26\pm8$\% respectively, if the GALEX detected objects are excluded or included. Even though these fractions are lower than in Nilsson et al.~(2009a), they are still $\sim 2.5$ times that what was reported at $z=3$ in Ouchi et al.~(2008). For a longer discussion on AGN fractions, see Nilsson \& M{\o}ller~(2011).
 The LAE AGN have total X-ray fluxes between $\log L_{0.5-10 \mathrm{keV}} = 43.3 - 45.5$~erg~s$^{-1}$, if at $z = 2.25$. 

\subsection{Spitzer detections}
The COSMOS field is covered by \emph{Spitzer} in the S-COSMOS survey (Sanders et al.~2007). The public data includes the four IRAC channels (3.6, 4.5, 5.8 and 8.0 $\mu$m), and the three MIPS bands (24, 70 and 160 $\mu$m). We will not discuss the 70 and 160~$\mu$m bands further as no detections were made in those bands. In the MIPS 24~$\mu$m band, both a full and a deep survey have been published, with the deep survey covering roughly 33\% of the original narrow-band image used for selection of candidates. The sensitivities of the \emph{Spitzer} data can be found in Table~\ref{spitzerdata}.
\begin{table}[t]
\begin{center}
\caption{S-COSMOS sensitivities from Sanders et al. (2007).  }
\begin{tabular}{@{}lcccccc}
\hline
\hline
Band  & CWL & FWHM & Sensitivity & \\
\hline
Ch1 & 3.58 & 0.75 & 0.0009 & \\
Ch2 & 4.50 & 1.02 & 0.0017 & \\
Ch3 & 5.80 & 1.43 & 0.0113 & \\
Ch4 & 8.00 & 2.91 & 0.0146 & \\
MIPS (deep) & 24.0 & 4.7 & 0.071 & \\
MIPS (shallow) & 24.0 & 4.7 & 0.15 & \\ 
\hline
\label{spitzerdata}
\end{tabular}
\end{center}
\begin{list}{}{}
\item[] Central wavelengths and full width half maximums are given in $\mu$m. Sensitivities are 5$\sigma$, and given in mJy. Only MIPS 24 $\mu$m values are given as our candidates have no detections in the longer wavelength bands.
\end{list}
\end{table}
We searched the public catalog (Sanders et al.~2007, v.~``GO2'', released May 2007) for sources associated with our candidates. In the IRAC bands, counterparts were searched for within a circle with 5 IRAC pixels ($3''$) radius around each candidate, and in the MIPS bands within a circle with 4 MIPS pixels ($4.8''$) radius. In total [56, 41, 25, 15, 20] candidate counterparts were found in the [Ch1, Ch2, Ch3, Ch4, MIPS-24] bands, see also Table~\ref{spitzdet} that includes all sub-sets of the candidates. 
\begin{table*}[t]
\begin{center}
\caption{Spitzer detections for all sub-sets of LAE candidates. }
\begin{tabular}{@{}lccccccccc}
\hline
\hline
Band  & Total sample & LAE best & GALEX only & AGN only & GALEX + AGN   \\
\hline
Ch1 & 56 (33\%)  & 36 (26\%) & 2 (29\%) & 13 (68\%)  & 5 (100\%)   \\
Ch2 & 41 (24\%)  & 24 (17\%) & 0 (0\%) & 12 (63\%)  & 5 (100\%)   \\
Ch3 & 25 (15\%)  & 8 (6\%)   & 0 (0\%) & 12 (63\%)  & 5 (100\%)   \\
Ch4 & 15 (9\%)  & 5 (4\%)   & 0 (0\%)  &  7  (37\%)  & 3 (60\%)  \\
MIPS & 20 (12\%)  & 8   (6\%)    & 0 (0\%)    &  7   (37\%) & 5  (100\%)  \\ 
\hline
\label{spitzdet}
\end{tabular}
\end{center}
\begin{list}{}{}
\item[] For sample definitions, see Table~\ref{tab:sampledef}. Percentages of each sample are given in parentheses.
\end{list}
\end{table*}
The random probability of finding an object within our search radii are 0.0096 in Ch1, resulting in 1.6 random false detections in the whole sample.
After visually inspecting the associations, two were excluded as random false detections and the list of Spitzer detected sources should, after removal of these two sources (detected in Ch1 and Ch2), be robust. The detection rate of our candidates is 33\% in Ch1 and 12\% in MIPS-24, which is a higher rate than found in surveys for LAEs at redshifts three and beyond (Lai et al.~2007, 2008) but in agreement with surveys at $z \sim 2.3$ (Colbert et al.~2006). The MIPS-24 detected candidates are by definition also ultra-luminous infrared galaxies (ULIRGs) due to their bright infrared fluxes. For a further discussion on these LAE ULIRGs and their evolution with redshift, see Nilsson \& M{\o}ller~(2009).

To test the public catalogues we performed aperture photometry on a number of objects in the catalogue. The catalogue fluxes were extracted in 1\farcs4 radius apertures (to be compared to 1\farcs5 radius apertures for the optical broad-band images), and aperture corrected to give the total flux of the object according to the aperture corrections listed in the release notes of the catalogue. The fluxes measured in the image were in reasonable agreement with the catalogue values, however, the errors measured were in the cases of Ch1 - Ch3 roughly two times those quoted in the catalogue. For Ch4 and MIPS the errors were also in agreement. Thus we multiply the published errors on our candidates in the Ch1 - Ch3 bands by two for the following analysis. There was a slight tendency of over-estimation of the fluxes in  Ch1, however, it was not significant enough to warrant a correction of the catalogue fluxes. It should be noted though, that the Ch1 fluxes may be over-estimated by up to 5\%. In the subsequent SED fitting, an extra error of $10$\% of the flux of each object was included, in order to account for systematic errors in the data reduction (see also Muzzin et al.~2009).

\section{SED fitting method}\label{sec:method}
The method used to fit the SEDs of the LAEs is an updated version of the code used in Nilsson et al.~(2007), called NisseFit. It is based on the stellar populations catalogue GALAXEV (Bruzual \& Charlot 2003) and the fitting is done using a Monte-Carlo Markov-Chain (MCMC) algorithm. The algorithm explores a multi-dimensional parameter space by stepping in a random fashion from one point $P_i$ in this space to another $P_{i+1}$. The step is random in the sense that the projected distance from point $P_i$ to $P_{i+1}$ along one of the parameter axes in the parameter space is a randomly chosen value ranging from 0 to the maximum step size, $\Delta P_{max}$.
The direction of the step along an axis is also random, with a probability of 50\% that the direction is along the positive or negative direction respectively. We impose boundary conditions for all parameters and ensure that no step leads to a value outside these boundaries.
The parameters and the boundary values we impose on them are listed in Table~\ref{tab:params}.
Once a new point $P_{i+1}$ is determined, the current model is advanced to the new point with a probability $p_{i\rightarrow i+1}$ which is given by the ratio of the two model probabilities given the data $D$,
\begin{equation}
p_{i\rightarrow i+1} = \frac{p(P_{i+1} | D)}{p(P_i|D)} = \mathrm{Max}(\exp{\left[(\chi^2_{i+1}-\chi^2_i)/T\right]},1),
\end{equation}\label{eq:chi}
where $\chi^2_{i+1}$ and $\chi^2_{i}$ are the reduced $\chi^2$ values for the two points in the parameter space and $T$ is a ``temperature'' parameter which can be used to adjust the convergence. In  our case $T=1$.
If the model is not advanced to the new point, the parameters remain $P_i$, so $P_{i+1}=P_i$.
This algorithm has several advantages over more common maximization algorithms (such as steepest descend, amoeba, etc.) in that it allows a full exploration of the parameter space. The result of our algorithm after $N$ iterations is a list of points $P_i, i=0..N$ that, if $N$ is sufficiently large, represents a good sampling of the underlying probability distribution function of the parameters. 

Please note carefully that, even though we refer to the quantity $\chi^2_r$ defined in Eq.~\ref{eq:chi} as the ``reduced chi-squared'' here and throughout this paper, the underlying modelling errors are not known to sufficient accuracy, compared to the relatively high measurement accuracy of the data, to allow the usual interpretation of a reduced chi-square value. Given the large number of photometric bands and the small measurement errors in many of these, no stellar population models are expected to give good fits to all the data in an absolute sense, and hence the $\chi^2_r$ values are not actually expected to be close to one. The term has been chosen here merely to reflect the fact that the ``goodness of fit'' is calculated in the same way as a reduced chi-squared. However, it should be interpreted as a relative merit value to compare different fits rather than an absolutely measurement of fit quality.

\subsection{Improvement on models}\label{sec:modelimprove}
\begin{figure}[t]
\begin{center}
\epsfig{file=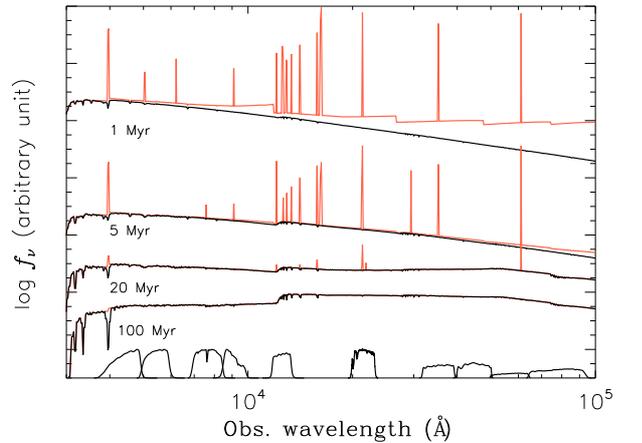,width=9.2cm}
\caption{Illustration of the added nebular emission. GALAXEV spectra at $z=2.25$ with ages 1, 5, 20, and 100 Myrs are shown with nebular emission lines (red) and without (black). The spectra are scaled to arbitrary units to illustrate the effect of age on the nebular emission. The filtercurves of the filters used in the fitting (Bj, Vj, $i^+$, $z^+$, J, K$_s$ and \emph{Spitzer} Ch1-4) are shown at the bottom of the figure. The nebular emission lines are added as delta functions at the respective wavelength of the emission lines, and the step-like appearance of the 1 Myr spectrum is due to the nebular continuum emission. At ages larger than 20 Myrs, the effect of the nebular add-on becomes negligible. }
\label{fig:nebex}
\end{center}
\end{figure}

Recent publications show that nebular emission lines can play an extremely important role in fitting SEDs (Zackrisson et al.~2008, Schaerer \& de Barros 2009, Raiter et al.~2010). In order to properly take into account this potentially large effect of gas emission in young, star-bursting galaxies, an add-on to the GALAXEV code was created. The nebular gas contributes with a continuum component and an emission-line component. The continuum addition was calculated from the Starburst99 models (Leitherer et al.~1999) with a Salpeter IMF with slope $\alpha = 2.35$ in the mass range $1 - 100$~M$_{\odot}$ (the Salpeter IMF is also used in the GALAXEV models). The emission-line strengths were calculated using the MAPPINGS photoionisation code (Kewley et al.~2011, in prep.). 
Both nebular continuum and emission-line strengths were calculated on a grid of ages (for the continuum in 36 steps between 1 and 100 Myrs and for the emission lines in six steps between 1 and 20 Myrs) and metallicities (same as available in GALAXEV, see below) and for each fit, the nearest values for these parameters were used. The emission-lines were scaled with the H$\alpha$ flux and the number of ionising photons in the galaxy spectra produced by GALAXEV. For an illustration of the added nebular emission, see Fig.~\ref{fig:nebex}.
In practice, the emission line strengths as well as the continuum contribution become negligible at ages $\gtrsim 20$~Myrs.

In this publication, we choose to fit only two single stellar populations (SSP) models, as the nebular emission add-on becomes too complicated with any other star formation history. For all models, dust is added according to Calzetti et al.~(2000). The parameter space allowed is detailed in Table~\ref{tab:params}.
\begin{table}[t]
\begin{center}
\caption{Parameter space allowed by SED fitting code.}
\begin{tabular}{@{}lccccccccc}
\hline
\hline
 & Min & Max & Remark  \\
\hline
Old pop. age & 0 Gyr & 2.7 Gyr &  \\
Young pop. age & 0 Gyr & 0.1 Gyr & $<$~Old age  \\
Mass frac. in young pop. & 0 & 1 &  \\
Dust A$_V$ & 0 mag & 5 mag &  \\
Metallicity $Z / Z_{\odot}$ & 0 & 1.0 & In steps, see text \\
Stellar Mass & --- & --- & Free parameter  \\
\hline
\label{tab:params}
\end{tabular}
\end{center}
\end{table} 
For the metallicity, the steps allowed are $Z / Z_{\odot} = 0.005, 0.02, 0.2, 0.4, 1.0$. Only fluxes in the bands from Bj to the $8.0\mu$m bands are used, as there are large uncertainties in the models at wavelengths below the Ly$\alpha$ line, and at restframe mid-IR wavelengths. The observed Ly$\alpha$ flux, as calculated from the narrow-band photometry, was subtracted from the Bj band before fitting to avoid uncertainties in the fitting of the Ly$\alpha$ line itself. Correspondingly, the Ly$\alpha$ line was also subtracted from the nebular emission add-on.

\begin{figure}[t]
\begin{center}
\epsfig{file=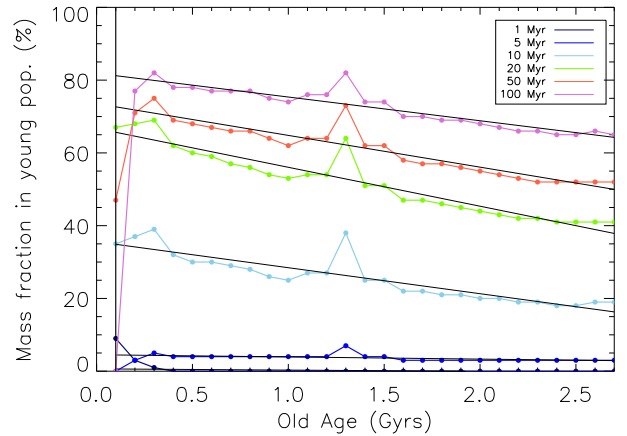,width=9.0cm}
\caption{Mass fractions of the young populations at which an old population is discernible from a young population. Galaxies with mass fractions fit \emph{below} these lines have two observable populations, for those with mass fractions fit \emph{above} the lines it is impossible to discern any older population. The limiting mass fraction where the older population becomes visible are shown with different colours, depending on the ages of the young population. At very young ages of the young population, all older populations are invisible. The solid vertical line marks the age limit of the young population in our fits. The solid horizontal lines are linear fits to the limiting mass fraction dependencies on the older population age. The peak at age $\sim 1.3$~Gyrs is a numerical artifact due to the limited age grid in the GALAXEV models. This does not affect our results as we use the solid, fitted lines for the measurements.}
\label{fig:massfrac}
\end{center}
\end{figure}

When the spectrum of a very young SSP is added to an older SSP spectrum, the older spectrum will need to comprise a very large mass fraction in order to be seen, as the young SSP will dominate the total spectrum. To test at what mass fractions the old population is actually seen, several young and old SSPs were generated and co-added with different mass fractions. To find an observability criterium for the older population, the colour in Vj$-$IRAC Ch1 was calculated. This was compared to the typical $1\sigma$ error bar in the photometry in these two bands. We calculated the colour significance according to:
\begin{equation}
col. signif. (mf) = \frac{(Vj-Ch1)_{mf} - (Vj-Ch1)_{mf=1.0}}{0.11}
\end{equation}
where $mf$ is the mass fraction. The number in the denominator is the typical error on the colour based on the data at hand. If this colour significance is more than $1\sigma$, the old population is considered to be observed. A plot of these limits can be found in Fig.~\ref{fig:massfrac}. Based on this figure, all objects with more than $90$\% of its data-points above the fitted lines are considered to be single young populations. 

\section{Results}\label{sec:results}
First we stacked the candidates in three different, light-weighted stacks for the fitting; the total sample of ``LAE best'' and GALEX-detected non-AGN LAEs, as well as the total sample divided into a ``red'' and a ``blue'' sample (with Vj$-i^+$ colours greater or less than zero respectively). These stacks are similar to those in Nilsson et al.~(2009a), but updated with spectroscopic results and new photometry from the UKIRT J and CFHT K$_s$ images. The magnitudes for the three samples can be found in Table~\ref{tab:sedmags}.
\begin{table}[t]
\begin{center}
\caption{Magnitudes for fitting of the stacked SEDs. }
\begin{tabular}{@{}lccccccccc}
\hline
\hline
  & Total & ``Red'' & ``Blue'' \\ 
\hline
\# & 122 & 80 & 42 \\
Bj & 25.29$\pm$0.05 & 25.22$\pm$0.04 & 25.36$\pm$0.12 \\
Vj & 25.02$\pm$0.06 & 24.92$\pm$0.11 & 25.22$\pm$ 0.09 \\
$i^+$ & 25.00$\pm$0.14 & 24.79$\pm$0.13 & 25.51$\pm$0.17 \\
$z^+$ & 24.92$\pm$0.22 & 24.61$\pm$0.15 & 25.46$\pm$0.23 \\
J & 24.02$\pm$0.24 & 23.63$\pm$0.22 & 24.47$\pm$0.45 \\
K$_s$ & 23.52$\pm$0.10 & 23.18$\pm$0.09 & 24.46$\pm$0.25 \\
Ch1 & $>23.46$ & 23.38$\pm$0.12 & $>23.46$ \\
Ch2 & $>23.46$ & 23.22$\pm$0.09 & $>23.46$ \\
Ch3 & $>20.70$& 22.79$\pm$0.13 & $>20.70$ \\
Ch4 & $>20.65$ & 22.75$\pm$0.19 & $>20.65$ \\
\hline
\label{tab:sedmags}
\end{tabular}
\end{center}
\begin{list}{}{}
\item[] First row gives number of galaxies in stack. Errors are the largest of either the background variance noise or the result of bootstrapping error calculation. 
\end{list}
\end{table}
Secondly, we also fit all individual objects with at least one detection in the Spitzer bands and/or in the K$_s$ band. This sample consists of a total of 58 objects (8 with only Spitzer detections, 19 with only a K$_s$ detection and 31 with detections in both Spitzer and K$_s$). Redshifts are set to 2.25 for all galaxies since varying the redshift within the range of the narrow-band filter does not change the fitting results significantly.

For the main results presented here, the Monte Carlo Markov Chain SED fitting code was run with $20$~$000$ iterations, and fitted the data with two single stellar populations. To find the probabilities in the parameter space, the first $2000$ runs were omitted to allow the $\chi^2$ to level out. 
The best fit parameters are calculated as the first and second moments of the distribution in each dimension, weighted by the $\chi^2_r$.
The full set of results of this run are found in Table~\ref{tab:fullresults}. When interpreting the dust parameter, A$_V$, it is important to keep in mind that this is a galaxy averaged dust absorption. Different parts of the galaxy will be observable at different wavelengths. The results for the dust parameter will be dominated by the absorption of the restframe UV part of the spectrum, see also section~\ref{sec:ulirgs}.

\begin{figure}[t]
\begin{center}
\epsfig{file=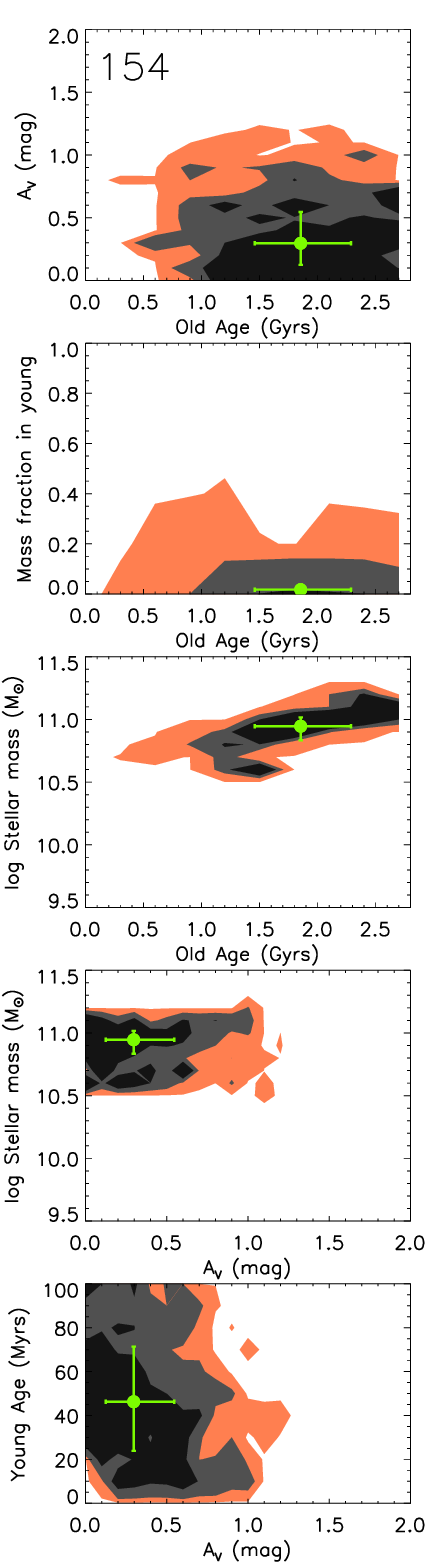,width=4.3cm}\epsfig{file=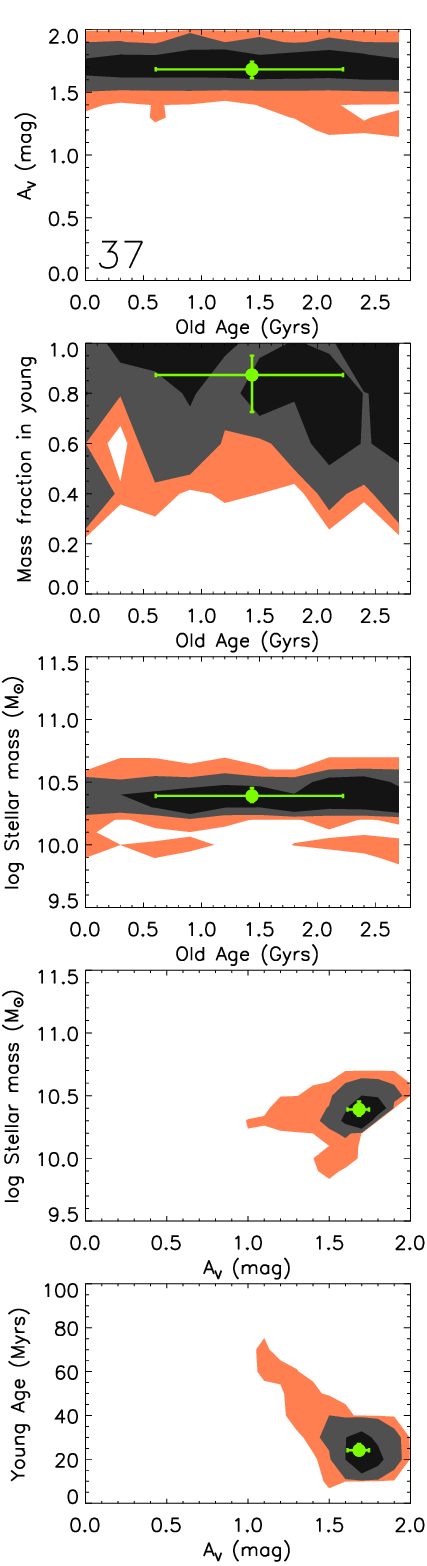,width=4.3cm}
\caption{Parameter dependencies between old/young population age, mass fraction, dust A$_V$ and stellar mass for LAE\_COSMOS\_154 (\emph{left}) and LAE\_COSMOS\_37 (\emph{right}). Contours mark 68\%, 95\% and 99.7\% probability. The green point marks the $\chi^2$-weighted best fit parameter, with the $1\sigma$ error bar as found in Table~\ref{tab:fullresults}. 
For the object in the left panels, the stellar mass is well constrained. The mass fraction in the young population is low, and hence the older population age is also well constrained. In this case, the younger population has a larger span of possible solutions, partially dependent on the A$_V$. This is an example of a galaxy for which the older population age is observed, as opposed to the object on the right. Here, the stellar mass and A$_V$ are well constrained, but the young population age is very young, and the mass fraction in the young population is allowed to be any value above $\sim 50$\%. Hence, the older population age is completely unconstrained. }
\label{fig:contour}
\end{center}
\end{figure}

When doing SED fitting, many of the fitted parameters can be degenerate with other parameters. In Fig.~\ref{fig:contour} we show, as an example, the contours of probability for the parameters of old population age, mass fraction in the young population, dust A$_V$, stellar mass, and young population age for two example objects (LAE\_COSMOS\_37 and 154), one that has a single young population and one that has a detectable older population (see also section~\ref{sec:indresults}).
Of the parameters fitted for, the stellar mass is the best constrained. Secondly, the dust A$_V$ is also well fit. Age of the old population, mass fractions and metallicities are the least constrained parameters. As can be seen in Fig.~\ref{fig:contour}, our calculation of the best fit value and its error bars is a good estimate of the distribution of parameters. No clear dependencies are seen. 
We will hereafter not discuss metallicities further, as they are in almost all cases completely unconstrained.

\subsection{Stack results}
The results for the stacked samples are shown in Table~\ref{tab:fullresults} (top). 
\begin{table*}[t]
\begin{center}
\caption{Full results of SED fitting run. }
\begin{tabular}{@{}lcccccccccc}
\hline
\hline
\multicolumn{10}{c}{Stacked samples} \\
\hline
Stack & Old Age & Young Age & Mass frac. & Dust A$_V$ & Metallicity & $\log$~M$_*$ & $\chi^2_r$ & SFR$_{corr}$ & AGN/GAL \\
 & Gyrs & Gyrs & (in young) &  &    $Z/Z_{\odot}$  & M$_{\odot}$ & &  M$_{\odot}$~yr$^{-1}$  & \\
\hline
Total non-K$_s$  & $1.13^{+0.85}_{-0.63}$ & $0.079^{+0.012}_{-0.020}$ & $0.43^{+0.26}_{-0.17}$ & $0.31^{+0.18}_{-0.18}$ & $0.4^{+0.0}_{-0.2}$ & $ 10.07^{+  0.15}_{-  0.12}$ & $  2.62^{+  0.86}_{-  0.59}$ &  ---    &    & \\
Red  & $1.31^{+0.75}_{-0.77}$ & $0.018^{+0.063}_{-0.004}$ & $0.55^{+0.25}_{-0.22}$ & $1.04^{+0.09}_{-0.35}$ & $0.4^{+0.0}_{-0.2}$ & $  9.87^{+  0.22}_{-  0.17}$ & $  4.67^{+  1.54}_{-  0.52}$ &   ---    &     & \\Blue  & $1.36^{+0.78}_{-0.81}$ & $0.046^{+0.019}_{-0.020}$ & $0.38^{+0.30}_{-0.19}$ & $0.11^{+0.14}_{-0.07}$ & $0.2^{+0.0}_{-0.18}$ & $  9.59^{+  0.24}_{-  0.26}$ & $  5.36^{+  0.83}_{-  0.64}$ &   ---    &     & \\
\hline
\multicolumn{10}{c}{Individual results} \\
\hline
1  & $0.76^{+1.16}_{-0.59}$ & $0.074^{+0.017}_{-0.030}$ & $0.53^{+0.33}_{-0.38}$ & $0.46^{+0.63}_{-0.26}$ & $0.4^{+0.6}_{-0.2}$ & $ 10.16^{+  0.23}_{-  0.09}$ & $  4.54^{+  0.98}_{-  0.81}$ & $   14.11\pm    0.89   $ &     & \\
10  & --- & $0.007^{+0.001}_{-0.001}$ & $>0.26$ & $2.30^{+0.04}_{-0.04}$ & $0.005^{+0.0}_{-0.0}$ & $ 10.57^{+  0.04}_{-  0.04}$ & $ 23.36^{+  0.98}_{-  0.60}$ &  ---  &     & \\
12  & --- & $0.022^{+0.005}_{-0.005}$ & $>0.52$ & $1.76^{+0.11}_{-0.10}$ & $0.005^{+0.0}_{-0.0}$ & $ 10.64^{+  0.06}_{-  0.05}$ & $  9.94^{+  2.02}_{-  1.08}$ & $   77.51\pm    5.16   $ &    G & \\
15  & $0.26^{+0.23}_{-0.04}$ & $0.077^{+0.015}_{-0.024}$ & $0.04^{+0.16}_{-0.03}$ & $0.12^{+0.18}_{-0.09}$ & $1.0^{+0.0}_{-0.6}$ & $ 10.60^{+  0.03}_{-  0.03}$ & $  5.72^{+  1.28}_{-  0.66}$ & $   11.64\pm    0.59   $ &     & \\
18  & $0.32^{+1.45}_{-0.30}$ & $0.017^{+0.007}_{-0.005}$ & $0.24^{+0.62}_{-0.20}$ & $2.13^{+0.16}_{-0.19}$ & $0.005^{+0.015}_{-0.0}$ & $ 10.74^{+  0.06}_{-  0.13}$ & $ 32.00^{+  1.97}_{-  0.98}$ &  ---  &     & \\
\hline
\label{tab:fullresults}
\end{tabular}
\end{center}
\begin{list}{}{}
\item[] Columns show $\chi^2_r$-weighted best fit parameters. Ages are in Gyrs and the mass fraction is the fraction of the total mass in the young population. Metallicity is fitted in steps, see the text for a description. The $\chi^2_r$ is the reduced chi-square. The eighth column gives the dust corrected UV star formation rates for the candidates with $\chi^2_r < 10$, see also sec.~\ref{sec:sfrcorr}. The final columns marks the entry with a G if the object is detected in the GALEX images. Full table is published as online material.
\end{list}
\end{table*}
A general comment for all results, both stacked and individual, is that the old population age is very hard to constrain and the error bars are typically several hundred Myrs, also in cases where there is a confirmed older population. There is a large consistency between the stack results. The old/young ages are similar between the stacks. As expected, the red versus blue stacks display diverging properties in dust and mass, with the red stack being dustier and more massive (A$_V$ = 1.04, M$_* = 7.3 \times 10^{9}$~M$_{\odot}$) than the blue stack (A$_V$ = 0.11, M$_* = 3.9 \times 10^8$~M$_{\odot}$). The total stack has values in between those of the red and blue stack results. The results of these three stacks are also well fitted, with $\chi^2_r < 4$.

\subsection{Interpretation of stack results}\label{sec:stackright}
When fitting SEDs at higher redshift, typically many objects are stacked in order to increase the signal-to-noise of the photometric measurements. But the question remains if the properties found fitting such a stack will be the average properties of the individual objects. To test this with our data, the 40 best fit objects were stacked, as well as the two sub-samples of single young population galaxies and those with an older population (see the sec.~\ref{sec:indresults}), and these stacks were fitted in an identical way as for the individual objects. The three stack were fitted with best $\chi^2_r =$~[3.78, 4.73, 4.41]. The results are presented in Table~\ref{tab:stackcomp}.
\begin{table}[t]
\begin{center}
\caption{Comparison of stack and individual fitting results. }
\begin{tabular}{@{}lccccccccc}
\hline
\hline
  & Age & A$_V$ & $\log$~M$_{\star}$ \\
  & (Gyrs) & (mag) & (M$_{\odot}$) \\ 
\hline
\multicolumn{4}{c}{All individually fit}\\
\hline
Mean of indiv. & $1.02\pm0.11$ & $0.76\pm0.04$ & $10.39\pm0.04$ \\
Stack & $0.44^{+0.40}_{-0.15}$ & $0.12^{+0.17}_{-0.08}$ & $10.23^{+0.09}_{-0.08}$ \\
$\sigma_{diff}$ & 1.40 & 3.66 & 1.69 \\
\hline
\multicolumn{4}{c}{Old pop. observed}\\
\hline
Mean of indiv. & $1.10\pm0.13$ & $0.63\pm0.04$ & $10.41\pm0.04$ \\
Stack & $0.56\pm0.16$ & $0.14^{+0.14}_{-0.08}$ & $10.42^{+0.06}_{-0.08}$ \\
$\sigma_{diff}$ & 2.62 & 3.37 & 0.13 \\
\hline
\multicolumn{4}{c}{Young pop. only observed}\\
\hline
Mean of indiv. & $0.012\pm0.001$ & $1.51\pm0.04$  & $10.30\pm0.05$ \\
Stack & $0.021\pm^{+0.042}_{-0.004}$ & $1.67\pm^{+0.13}_{-0.11}$  & $10.45^{+0.05}_{-0.04}$  \\
$\sigma_{diff}$ & 2.18 & 1.37 & 2.37 \\
\hline
\label{tab:stackcomp}
\end{tabular}
\end{center}
\begin{list}{}{}
\item[] Table gives the best fit parameters for the mean and standard deviation of the mean of individually fit objects and the $\chi^2$-weighted parameters of the light stack of the same objects. The results at the top of the table are for the whole sample of 40 best fit objects, whereas the bottom gives the results for the two sub-samples of galaxies with only a single young population observed (6) or with an older population observed (34). The trends of ages and A$_V$s decreasing when fitting a stack are obvious in the total and old sample.
\end{list}
\end{table}

The values from the mean of the individual fits and those from the stack are not in great agreement as seen in Table~\ref{tab:stackcomp}. The stellar masses give robust results in all cases, but the ages are about a factor of two younger in the stacks than the mean of the individual fits for the total sample and the old population sub-sample. Further, the dust extinction is lower by about a factor of five in the stacks than in the mean of the individual fits. We interpret these results as being due to a smearing of the UV slopes of the SEDs of the galaxies in the stacking. It is clear that results from stacking of objects, especially those of the dust extinction, should be interpreted with caution and may be considered lower limits.

\subsection{Individual results}\label{sec:indresults}
\begin{figure}[t]
\begin{center}
\epsfig{file=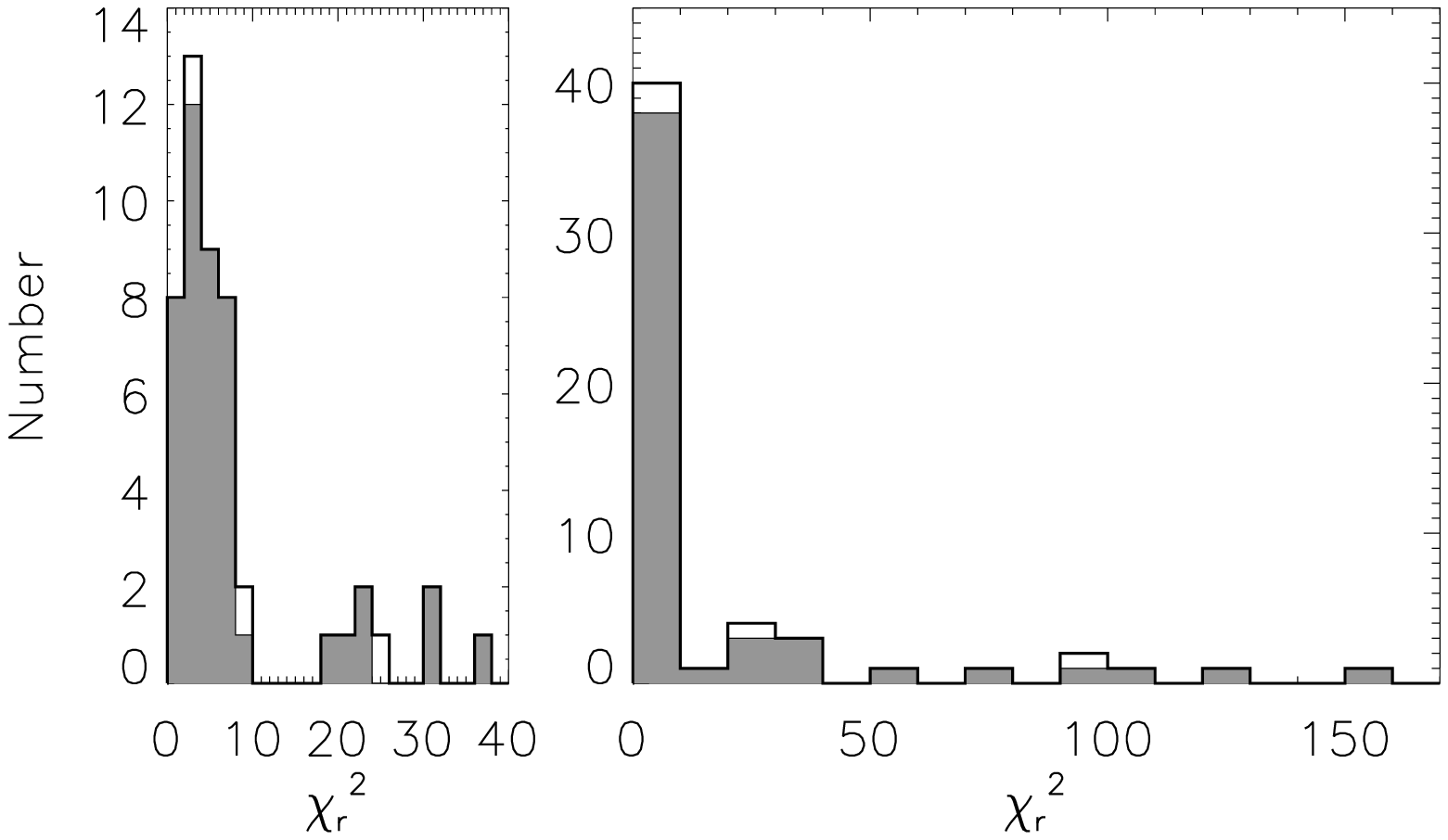,width=9.0cm}
\caption{Histogram of $\chi^2_r$ of non-AGN selected candidates. Right panel shows complete distribution with a bin size of 10, left panel a zoom on the range $0 - 40$~$ \chi^2_r$ with a bin size of 2. 
Filled histogram is for candidates that are not detected in the GALEX data.}
\label{fig:chihist}
\end{center}
\end{figure}

\begin{figure*}[t]
\begin{center}
\epsfig{file=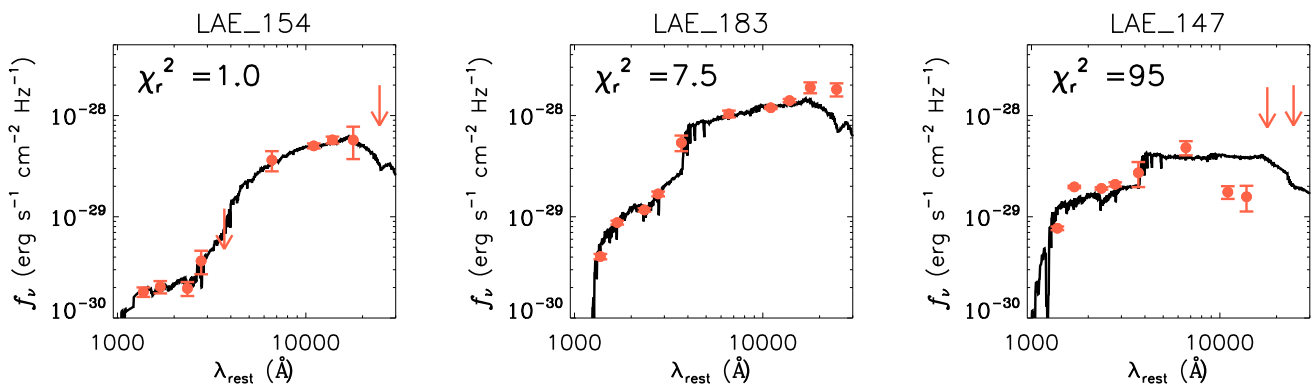,width=18.0cm}
\caption{Illustration of the goodness of fit with various $\chi^2_r$ from three candidates. The $\chi^2_r$ are 1, 8 and 95 respectively from left to right in the figure. As can be seen, also fits with $\chi^2_r \sim 10$ are reasonably good fits.}
\label{fig:testchi}
\end{center}
\end{figure*}

The the best fit parameter values for all individually fitted objects are found in Table~\ref{tab:fullresults} (bottom). In this section we remark on the general properties of all these objects, and on one special case. Of the, in total 58, objects fitted, four are detected in the GALEX data, of which three are spectroscopically confirmed. In Fig.~\ref{fig:chihist} the histogram of the best fit $\chi^2_r$ are shown. 
An obvious break is seen at $\chi^2_r \approx 10$ in the distribution. In Fig.~\ref{fig:testchi}, three examples of SEDs and best fit spectra with very different $\chi^2_r$ are shown. 
As explained in sec.~\ref{sec:method}, the $\chi^2_r$ is not expected to be unity and it is clear that fits with $\chi^2_r \approx 10$ still appear to be very good fits, as seen in the example of LAE\_COSMOS\_183, Fig.~\ref{fig:testchi}. We thus choose to define objects as ``well fitted'' if its calculated merit value ($\chi^2_r$) is less than ten.
Of the 58 fitted candidates, 40 have $\chi^2_r < 10$. Two of these candidates are GALEX-detected, both of which are spectroscopically confirmed.
For the remaining candidates with very bad $\chi^2_r$, about half have Spitzer data points lying significantly off from the other SED points (see e.g. LAE\_COSMOS\_147, Fig.~\ref{fig:testchi}), and the other half have unusual SEDs, possibly consistent with lower redshift. Because of the reasons mentioned above, the results for the galaxies with bad $\chi^2_r$ are not reliable and we will concentrate on the 40 candidates with $\chi^2_r < 10$ in the further analysis.

\begin{figure}[!ht]
\begin{center}
\epsfig{file=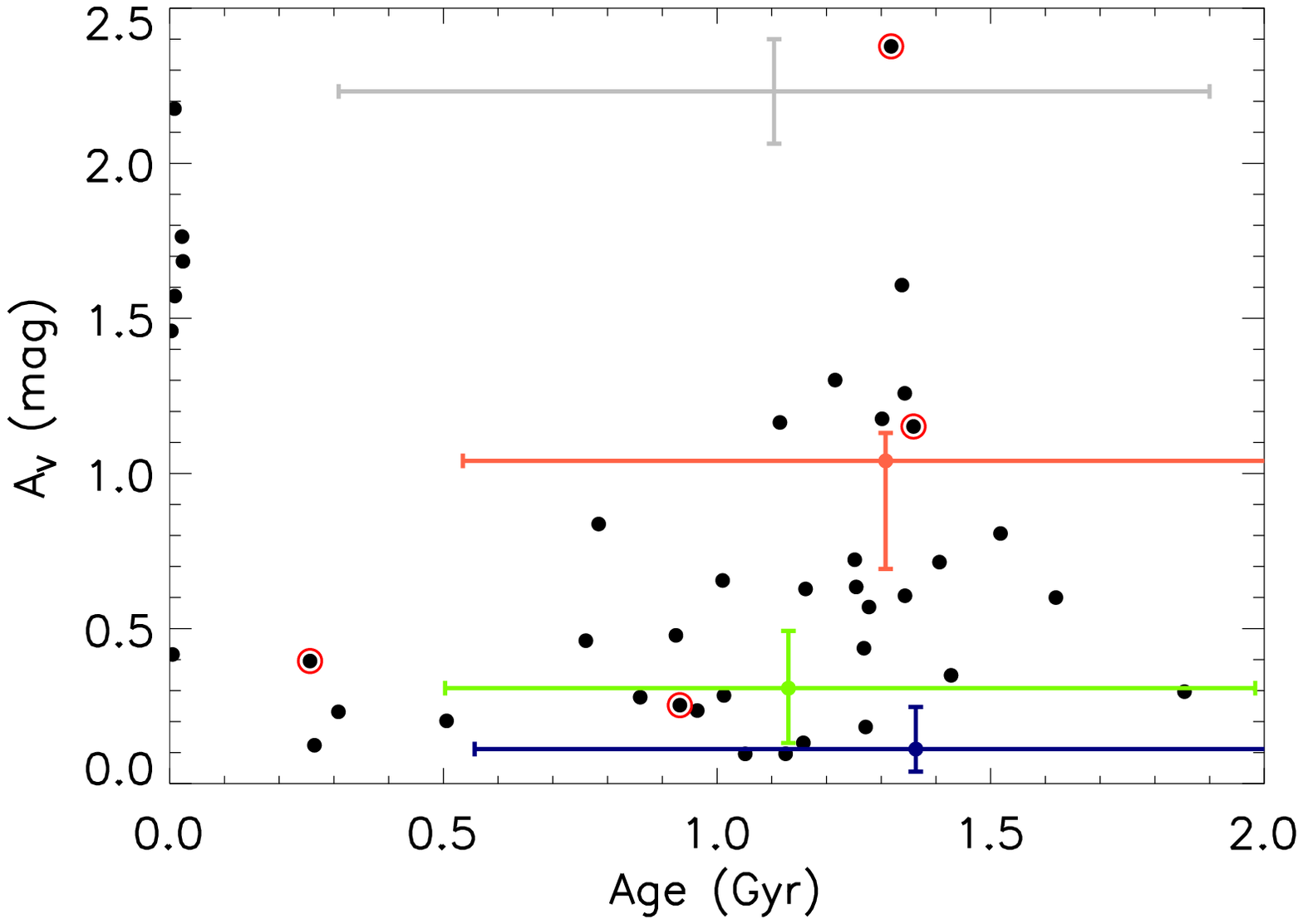,width=9cm}
\epsfig{file=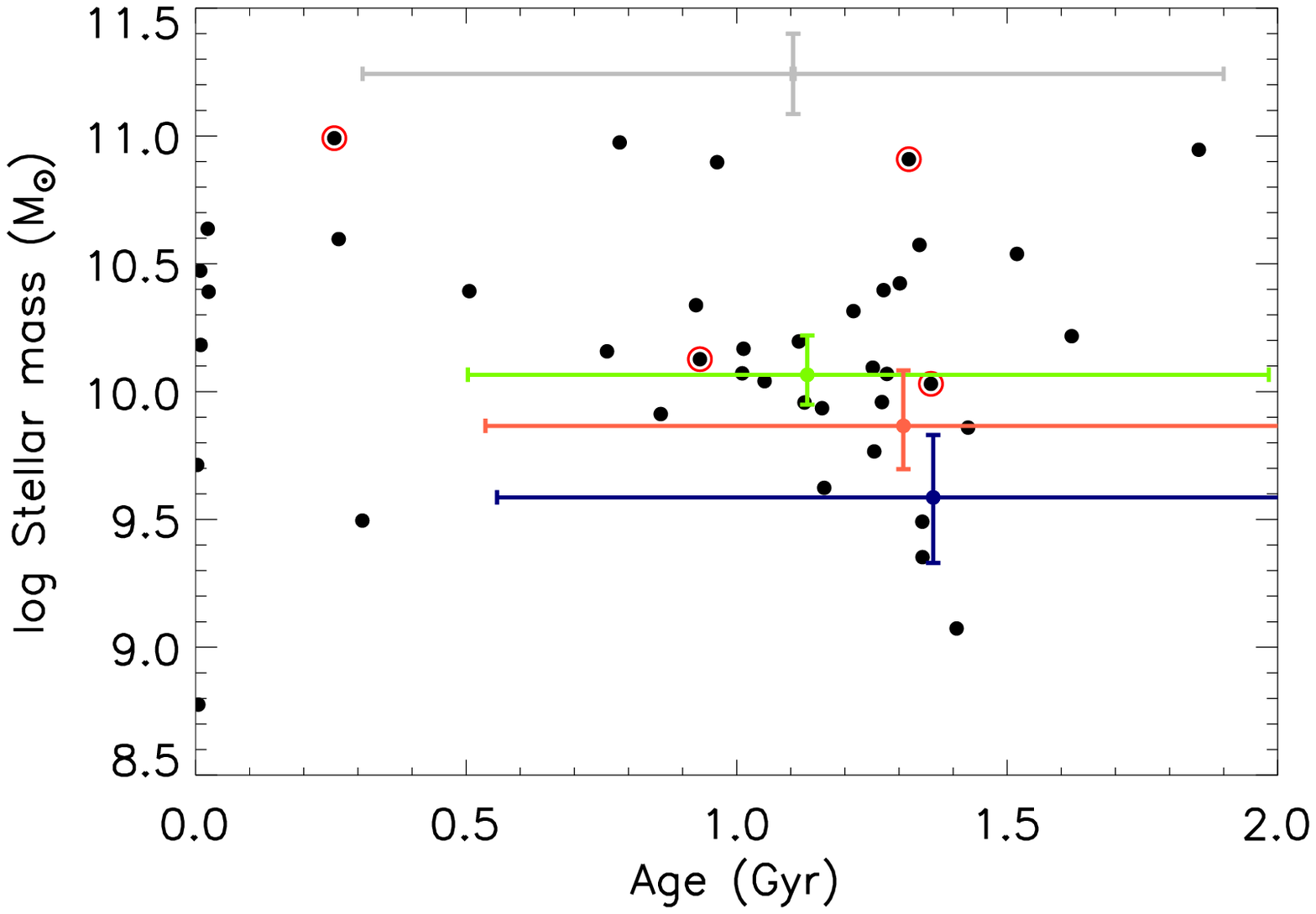,width=9cm}
\epsfig{file=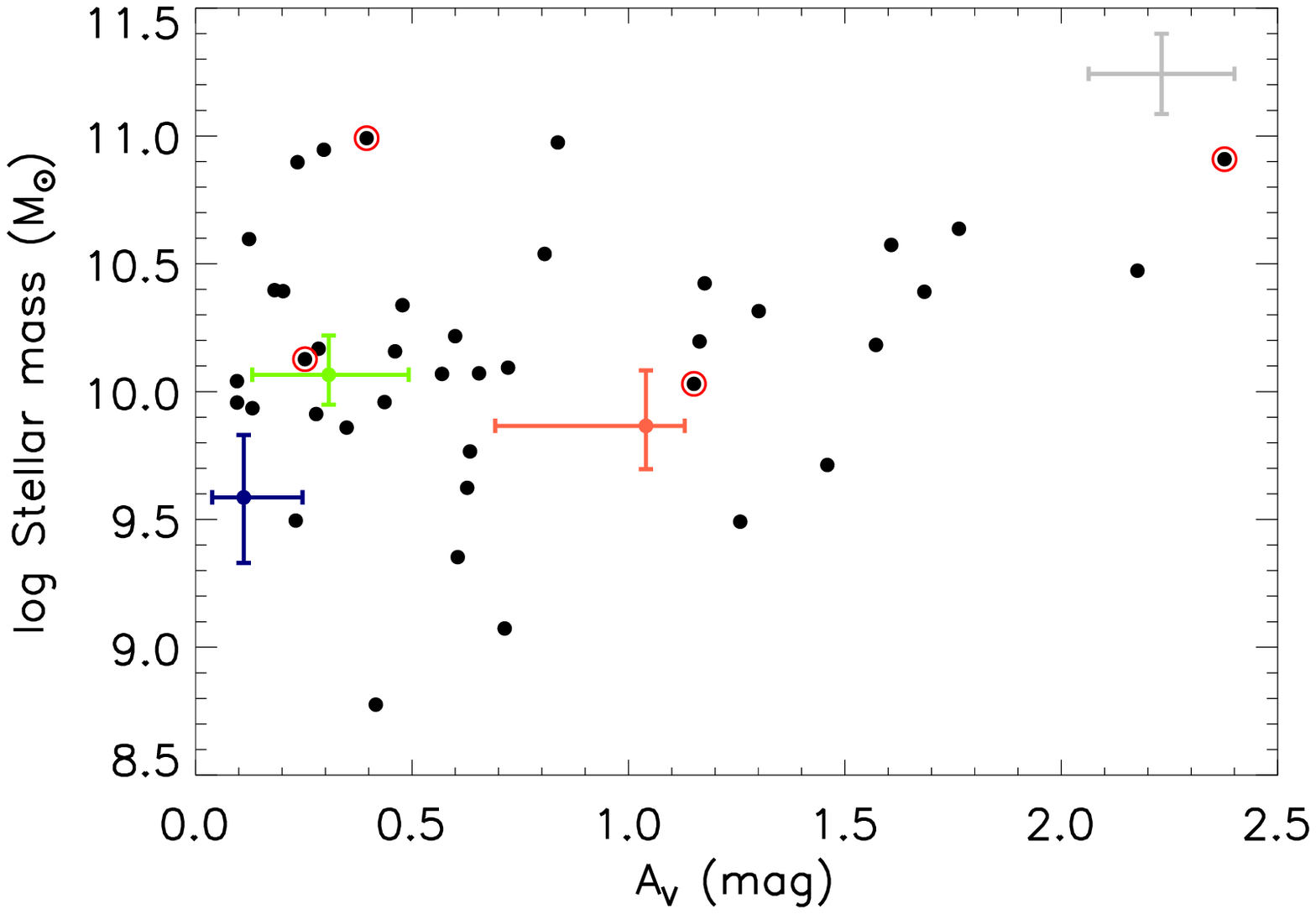,width=9cm}
\caption{Results of 40 objects with $\chi^2_r < 10$ (black points). The coloured data points represent the stacked SEDs, with green as the total sample, red and blue points are the red and blue sub-samples. Typical error bars are marked in grey. In the two top panels the error bars in the ``Age'' direction refer to those with ages above 100~Myrs. For the single young population objects, the typical error in age is $\sim 15$~Myrs. LAE ULIRGs (from Nilsson \& M{\o}ller~2009) are marked with a red ring. }
\label{fig:results}
\end{center}
\end{figure}

For all galaxies, the test shown in Fig.~\ref{fig:massfrac} was performed in order to determine if a galaxy was best described by two stellar populations, or could be best fitted with a single young population of stars. Hereafter, the ``Age'' of a galaxy is considered to be the age of the young population, if no older population can be seen, and the age of the older population, if this is considered observed (cf.~Fig.~\ref{fig:contour}). The results from the SED fits can be found in Fig.~\ref{fig:results}. 
Of the galaxies $\sim 15$\% exhibit single young populations ($15^{+9}_{-6}$\%). Conversely, $\sim 85$\% can be seen to host at least one population of older stars. This is similar to the results of Guaita et al.~(2010). If the stellar mass in the two sub-samples is summed up, $12$\% of the total stellar mass is in the young galaxies, and $88$\% is in the old. Hence, the stellar mass per galaxy is on average slightly higher in those galaxies with an older population. The older ages of these multiple population galaxies range between $0.5 - 2$~Gyr with a mean of $1.10\pm0.13$~Gyrs, but are relatively unconstrained. The ages of the single, young population galaxies are $12\pm1$~Myrs. The dust A$_V$ values for the whole sample range between $0.0 - 2.5$~mag, with mean A$_V = 0.63\pm0.04$~mag and $1.51\pm0.05$~mag for the old and young populations. The stellar masses are in the range $8.5 < \log \mathrm{M}_* < 11$~M$_{\odot}$, with, for the old/young populations, $\log \mathrm{M}_* = 10.4\pm0.04$~/~$10.3\pm0.05$~M$_{\odot}$. In Table~\ref{tab:fullresults}, the lower limits are given on the mass fractions in the young population in those galaxies with a single young population observed.

\subsubsection{LAE ULIRGs}\label{sec:ulirgs}
It was shown in Nilsson \& M{\o}ller~(2009) that some of these candidates have infrared fluxes consistent with being ultra luminous infrared galaxies (ULIRGs). We now ask the question if these galaxies are located in a certain part of the parameter space fitted, or if they are evenly distributed. Of the non-AGN ULIRGs in the sample, four were fitted with good $\chi^2_r$ and these are marked with a red ring in Fig.~\ref{fig:results}. They appear evenly distributed throughout the diagrams. Further, they are generally above the average in stellar mass and all contain a significant older stellar population, but are very evenly distributed in the extinction parameter. The extinction in the UV/optical ($A_V$) was derived from the ratio of UV to infrared fluxes in Nilsson \& M{\o}ller~(2009), and ranged between $4.8 - 7.2$~magnitudes for the four that have been fitted here. Clearly, the two methods, of determining the dust extinction primarily based on the UV slope or including infrared data points, reveal vastly different extinction values. The most likely explanation for this discrepancy is that they are due to dust geometry effects, as both methods depend differently on the assumption that the dust is distributed in a homogeneous screen in front of each object.

\subsubsection{LAE\_COSMOS\_94}\label{sec:lae94}
\begin{figure}[t]
\begin{center}
\epsfig{file=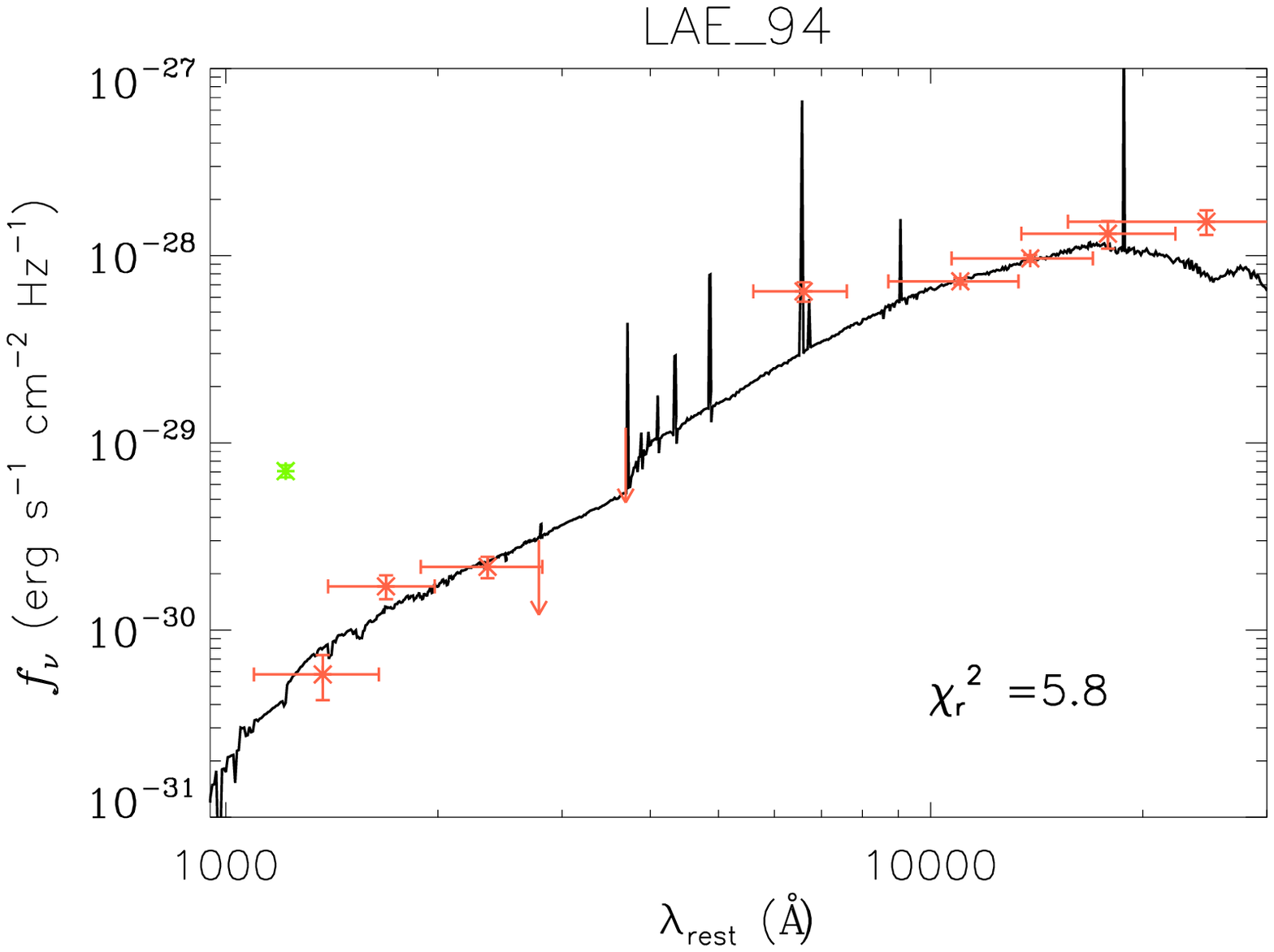,width=9.0cm}
\caption{SED of LAE\_COSMOS\_94. It is very faint in the broad-bands, although with a large Ly$\alpha$ flux, and a very red SED in the \emph{Spitzer} bands. Red points are observed data-points, arrows mark upper limits. The solid line shows the best fit spectrum. The green data point is the narrow-band flux (Ly$\alpha$) for reference.}
\label{fig:lae94}
\end{center}
\end{figure}

One very interesting example of the LAE ULIRGs is LAE\_COSMOS\_94, fitted with a best $\chi^2_r = 5.8$. This object, which is confirmed to be a Ly$\alpha$ emitter by the spectroscopy, is the one in our survey with by far the highest equivalent width ($EW_0 = 768$~{\AA}). It has a Ly$\alpha$ magnitude of $24.33 \pm 0.10$ but is very faint in the optical broad-bands with magnitudes $\sim 26$ and is even undetected in the $z^+$ and J bands, but then shows a very red SED in the K$_s$ and \emph{Spitzer} bands, see Fig.~\ref{fig:lae94}. 
The best fit properties of this galaxy are age~$= 1.32^{+0.80}_{-0.73}$~Gyr, A$_V =2.38\pm0.13$, and $\log$~M$_* = 10.91^{+0.26}_{-0.22}$~M$_{\odot}$. These properties are very similar to the average properties of all the candidates, except it has the largest A$_V$ of all.

Very high Ly$\alpha$ EWs have been proposed to be signs of Population III stars (Schaerer~2003, Tumlinson et al.~2003), whereas LAE\_COSMOS\_94 is obviously not a Population III object. Very high EWs could also appear in dusty objects, where the dust geometry allows Ly$\alpha$ photons to escape (cf.~Finkelstein et al.~2008). This object also appears to be quite extended in the narrow-band image, with a PSF-subtracted FWHM of 2\farcs04, corresponding to 16~kpc at $z = 2.25$. The PSF-subtracted FWHM in the Bj/$r^+$ bands are 1\farcs28/1\farcs04, or 10/8 kpc. The narrow-band, i.e. Ly$\alpha$ emission, is clearly more extended than the broad-band component, as has previously been found for other LAEs at both low redshift ({\"O}stlin et al.~2009) and at high redshift (M{\o}ller \& Warren 1998, Fynbo et al.~2001, Finkelstein et al.~2010). This is further evidence that the high EW in this object is driven by resonant scattering, as the Ly$\alpha$ emission is expected to be more extended than the continuum emission in those cases ({\"O}stlin et al.~2009).

\subsubsection{Correlations in the data}
To test if any correlations exist between the typical parameters stellar mass, dust A$_V$ and age in the full sample, a Spearman rank test was performed. The Spearman rank test allows to test the existence of any monotonous relationships in the data. The results of the Spearman rank test can be found in Table~\ref{tab:spearman}.
\begin{table}[t]
\begin{center}
\caption{Results for the Spearman rank test on several parameter correlations. }
\begin{tabular}{@{}lccccccccc}
\hline
\hline
   Parameters & Spearman rank $\rho$ & Correlation prob.  \\
\hline
  mass $-$ A$_V$ & $0.17$ & 86\%  \\ 
  age $-$ A$_V$ & $0.03$ & 68\%  \\ 
  mass $-$ age & $-0.10$ & 86\%  \\ 
  Ly$\alpha$ Flux $-$  mass & $-0.20$ & 86\%    \\
  Ly$\alpha$ Flux $-$  A$_V$ & $-0.23$ & 91.4\%    \\
  Ly$\alpha$ EW $-$  mass & $-0.08$ & 68\%    \\
  Ly$\alpha$ EW $-$  A$_V$ & $-0.16$ & 78\%    \\
\hline
\label{tab:spearman}
\end{tabular}
\end{center}
\begin{list}{}{}
\item[] The Spearman rank number is a measure of the correlation in the data. Points along a perfectly monotonic function would have a Spearman rank of $\pm1$. A positive rank number means a positive correlation between the parameters, and a negative number an anti-correlation between the parameters. 
\end{list}
\end{table}
No significant correlations are seen.

\begin{figure*}[t]
\begin{center}
\epsfig{file=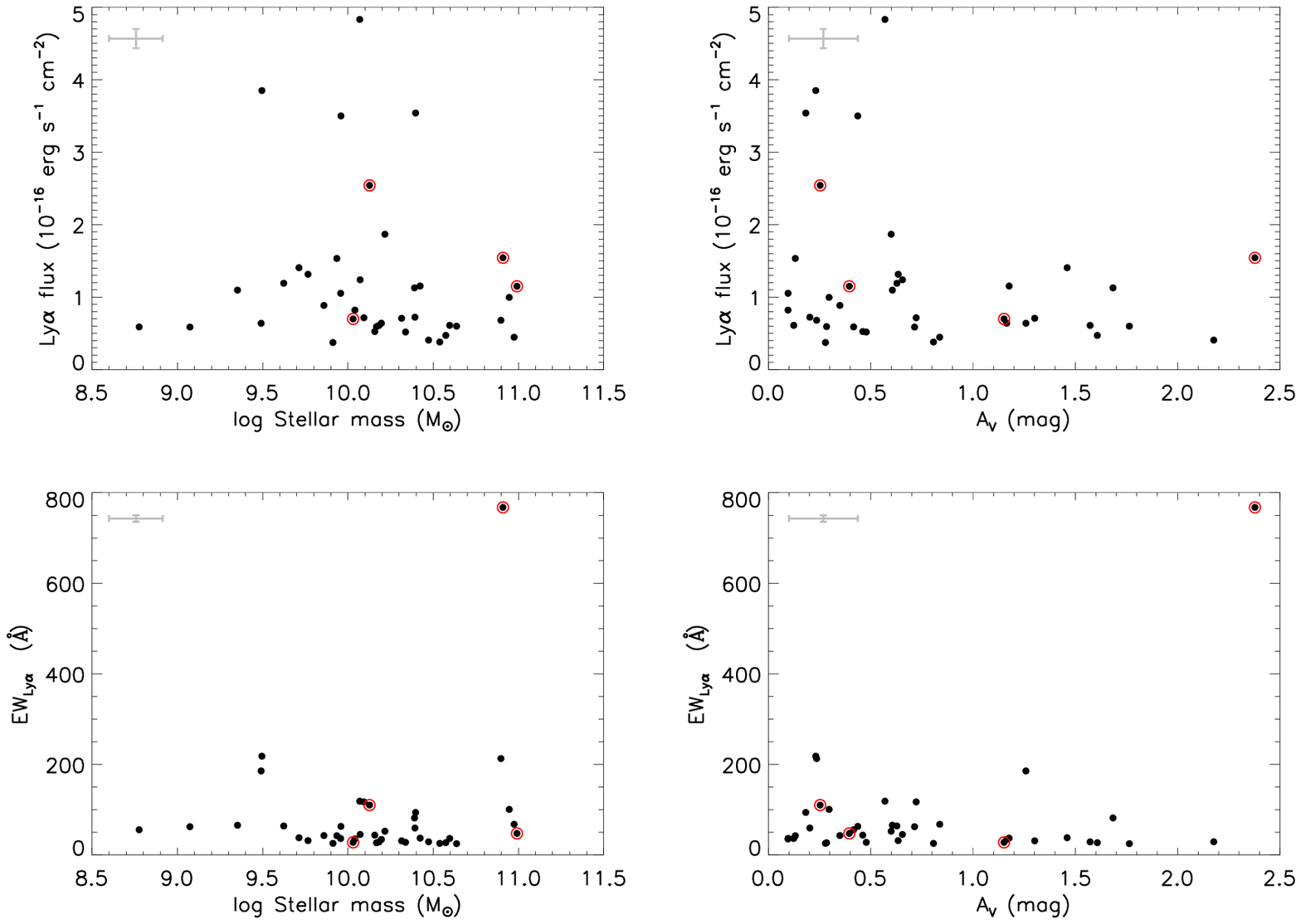,width=18.0cm}
\caption{ Dependencies between Ly$\alpha$ flux (upper panels) and EW (lower panels) and best fitted stellar masses (left panels) and dust A$_V$ (right panels). Typical error bars are marked in grey. ULIRGs are marked with a red ring. The very high EW object in the lower panels is LAE\_COSMOS\_94, see sec.~\ref{sec:lae94}. No clear dependencies are seen, besides an apparent lack of objects at large Ly$\alpha$ fluxes and large extinctions.}
\label{fig:lyased}
\end{center}
\end{figure*}

With the data at hand, it is also possible to study dependencies between the Ly$\alpha$ fluxes and equivalent widths and the parameters in the SED fit. In Fig.~\ref{fig:lyased}, the Ly$\alpha$ fluxes and EWs of the galaxies are shown as function of stellar masses and dust A$_V$.
In the top panels in this figure, the Ly$\alpha$ flux is shown as a function of stellar mass and dust parameter. A lack of objects with large Ly$\alpha$ flux and large dust A$_V$ is seen in the top right panel, although it is not statistically significant. In the lower panels of Fig.~\ref{fig:lyased}, the Ly$\alpha$ restframe EW is shown as a function of stellar mass and dust. The object with extremely high EW is LAE\_COSMOS\_94, see also sec.~\ref{sec:lae94}. No correlation between stellar mass and EW is seen, as proposed in Pentericci et al.~(2009). The Spearman rank test was applied also to these data-sets, revealing only a weakly significant correlation between Ly$\alpha$ flux and A$_V$, see Table~\ref{tab:spearman}.

\subsection{Dust corrected star formation rates}\label{sec:sfrcorr}
When the dust absorption for each galaxy is determined, the UV spectrum can be corrected and the intrinsic UV flux retrieved. From the intrinsic fluxes, dust corrected star formation rates can be derived according to the conversion of Kennicutt (1998). The uncorrected SFRs were presented in Nilsson et al. (2009a), where the UV-derived SFRs of all candidates were found to be in the range $1 - 40$~M$_{\odot}$~yr$^{-1}$. For the galaxies with good fits to the SED, the Vj band (used for the SFR calculation) can be corrected with the best fitted A$_V$. Within this sub-sample of 40 galaxies, the maximum SFR$_{UV}$ before correction was $15$~M$_{\odot}$~yr$^{-1}$ and the median $\sim 8$~M$_{\odot}$~yr$^{-1}$. After correction, the range of SFRs in this sub-sample is $4 - 120$~M$_{\odot}$~yr$^{-1}$, with a median of $16$~M$_{\odot}$~yr$^{-1}$. These values are more similar to $z \approx 2$ galaxies (such as BX/BM, BzK and DRG galaxies), although still in the lower two-thirds of the range covered by those galaxies (Reddy et al. 2005). The dust corrected star formation rates of the well fit objects are given in Table~\ref{tab:fullresults}.

\section{Redshift evolution in properties}\label{sec:discussion}
\begin{figure}[!t]
\begin{center}
\epsfig{file=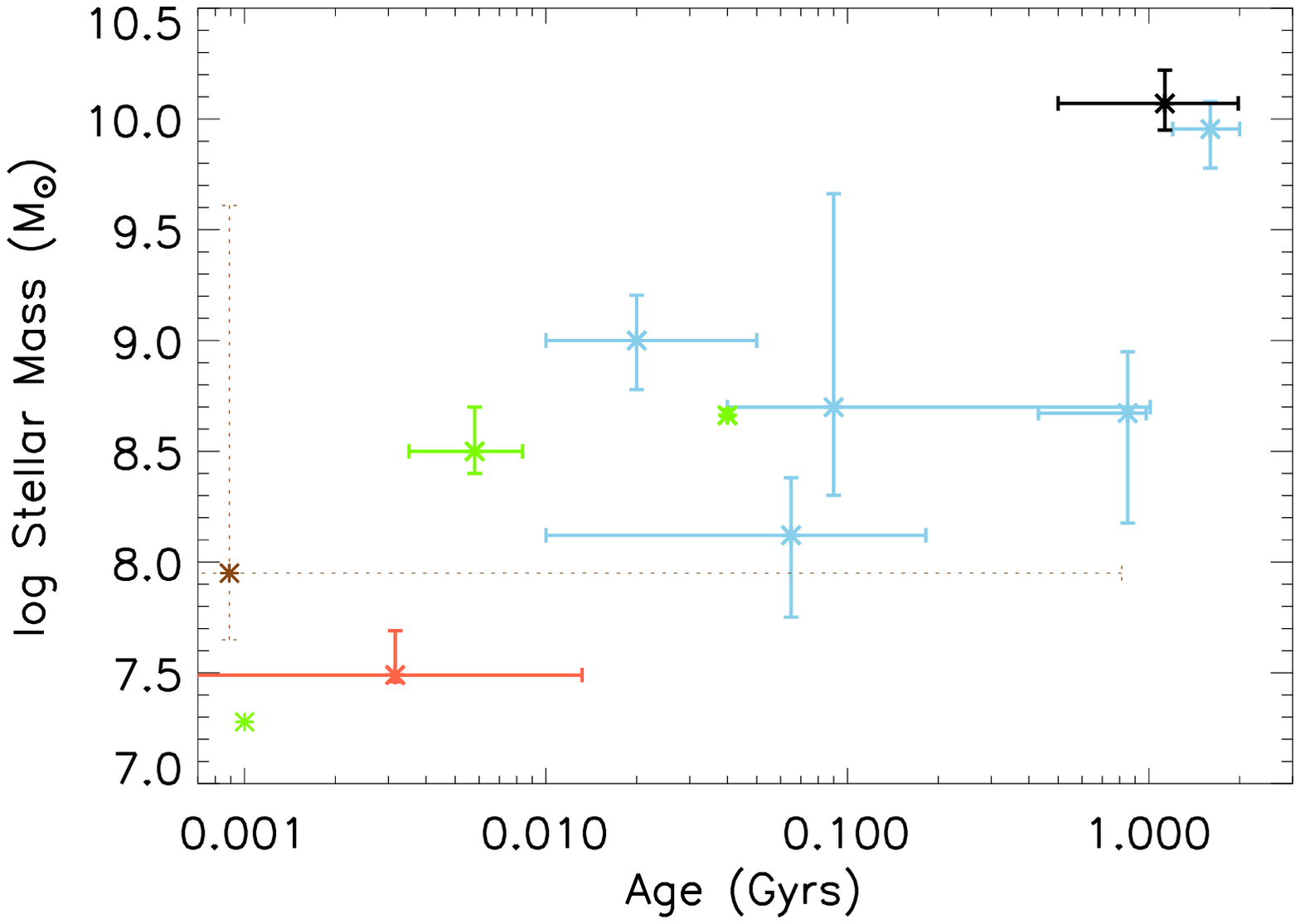,width=8.4cm}
\caption{ SED properties derived for stacked LAEs in several publications at different redshifts; at $z \sim 3$ with light blue points from Gawiser et al. (2006, 2007), Nilsson et al. (2007), Lai et al. (2008) and Ono et al.~(2010a). Green points are at $z = 3.7 - 4.5$ from Finkelstein et al. (2007) and Ono et al.~(2010a). The red point is at $z \sim 5.7$ from Ono et al.~(2010b) and the brown point is from Ono et al.~(2010b) for $z = 6.5$ LAEs.
The black star is the result of the fit of the total stack of LAEs (122 objects) at $z=2.25$. Points without error bars were published without these values.}
\label{fig:properties1}
\end{center}
\end{figure}

We now wish to study how the properties of LAEs have changed with redshift, but results from SED fits of stacked photometry should be interpreted with caution, as noted in sec.~\ref{sec:stackright}. Stellar masses appear to be robust to stacking, but in this case there is a bias between stacks and individual fits in that individual fits are typically only done on the brighter galaxies, which will then be biased towards more massive objects. Hence, to compare stellar masses, it is more advantageous to study stacked results, as these will be less sensitive to flux limitations. Further, even though the age of a stack was less than the mean age from the stacked objects fitted individually, the results were still relatively consistent with those from fitting individual objects. In Fig.~\ref{fig:properties1} we show the stellar masses and ages for stacks reported in several previous papers.
In this figure, a clear correlation between age and stellar mass is seen, as confirmed by a Spearman rank test assigning a 99.94\% probability to it. It also appears the LAEs typically get older and more massive with lower redshift. This trend will not be affected by the potential bias in younger ages found when stacking, as such a bias would be likely identical at all redshifts, and would thus only shift the relation in the age direction. Note though that there is an intrinsic degeneracy between ages and masses due to increasing mass-to-light ratios in older galaxies. 

We further see that the $z=2.3$ LAEs are among the most massive LAEs so far found. To rule out a possible bias resulting from continuum detection levels, we search the literature, from where the data-points in Fig.~\ref{fig:properties1} are taken, for information. For each redshift, the continuum limit as close to restframe $\sim 4100$~{\AA} as possible was noted. This corresponds to a location just redward of the Balmer break which should ensure a less dust-sensitive/more mass-sensitive data-point. The stacked continuum limit was then calculated from the $3\sigma$ limit in the images used in each case, corrected for the number of objects in each stack. Finally, these limits were corrected for redshift dimming by converting to absolute magnitudes. Comparing these absolute, stacked continuum limits for the eleven measurements in Fig.~\ref{fig:properties1} reveals that the stacked limit for the sample here is the faintest of all, with an absolute stacked magnitude limit of -18.1 ($3\sigma$, AB) and with the other ten results having varying continuum limits between -18.2$\, - \, $-20.7. Since the stellar mass in principle is directly proportional to the flux in the red part of the SED (excluding small dust effects), it is clear that any existing, smaller stellar mass objects would have been detected in our stack, and that the conclusion that $z=2.3$ LAEs are more massive than those at higher redshifts is robust. Any bias, if present, would lead to lower average stellar masses derived for the $z=2.3$ LAEs than for higher redshift LAEs, and therefore our conclusion of larger stellar masses at lower redshift is strengthened.

\begin{figure}[!t]
\begin{center}
\epsfig{file=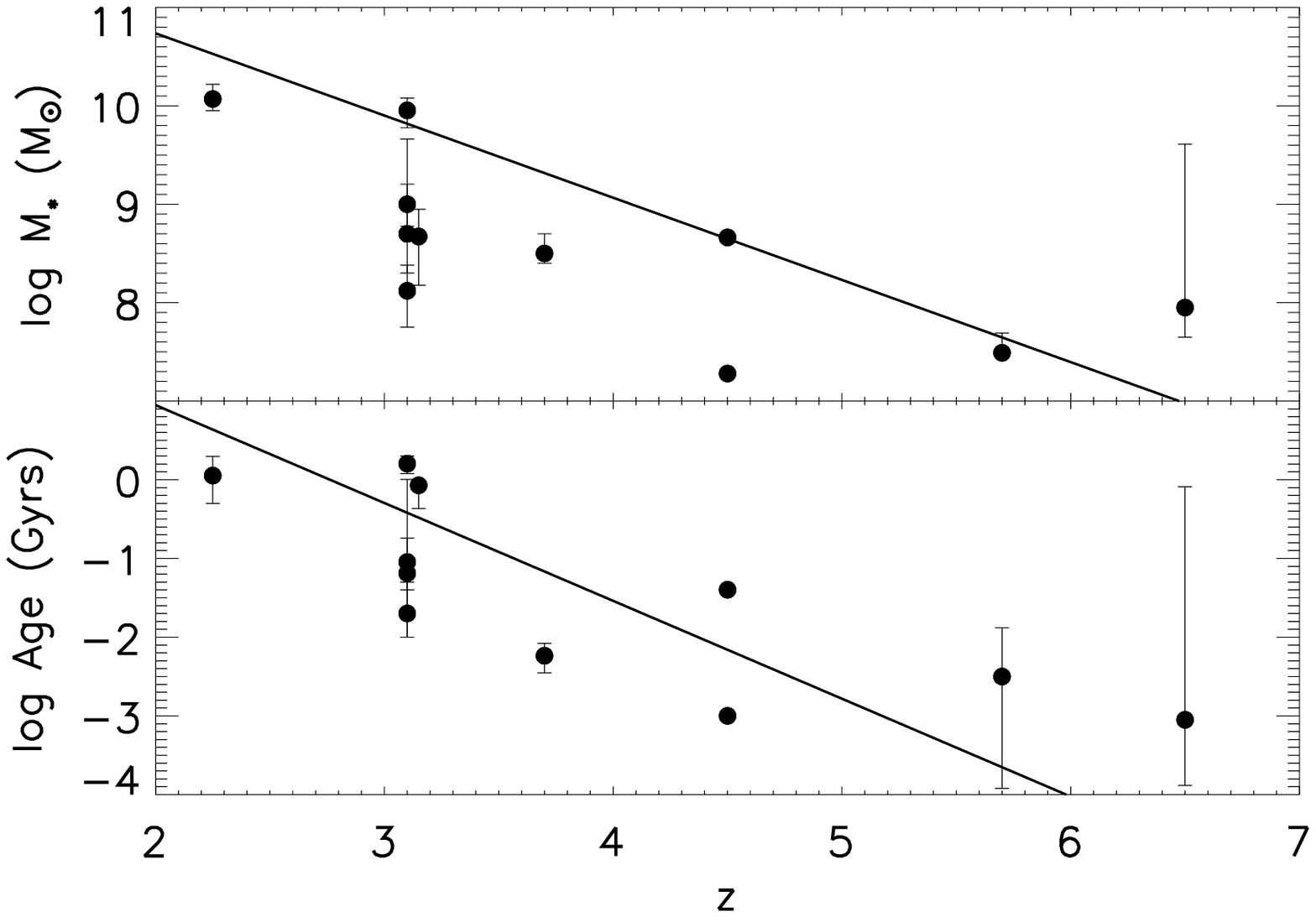,width=8.4cm}
\caption{Panels show the SED derived ages and masses as a function of redshift from the different surveys mentioned in the caption for Fig.~\ref{fig:properties1}. The solid lines are best fit linear relations. }
\label{fig:agemasscorr}
\end{center}
\end{figure}

To further study the apparent redshift dependence on the stellar mass and the ages, we plot the data-points from Fig.~\ref{fig:properties1} separately for stellar mass and age as a function of redshift in Fig.~\ref{fig:agemasscorr}. 
The Spearman rank probability of a correlation in the data is 99.92\% for age and redshift and 99.45\% for stellar mass and redshift. The solid line in the figure represents linear fits to the data, with the following parameters.
\begin{equation}
\log \mathrm{Age} \, \mathrm{(Gyrs)} = 3.43 - 1.24 \times z
\end{equation}
\begin{equation}
\log \mathrm{M}_* \, \mathrm{(M}_{\odot}\mathrm{)} = 12.41 - 0.84 \times z
\end{equation}

\begin{figure}[!t]
\begin{center}
\epsfig{file=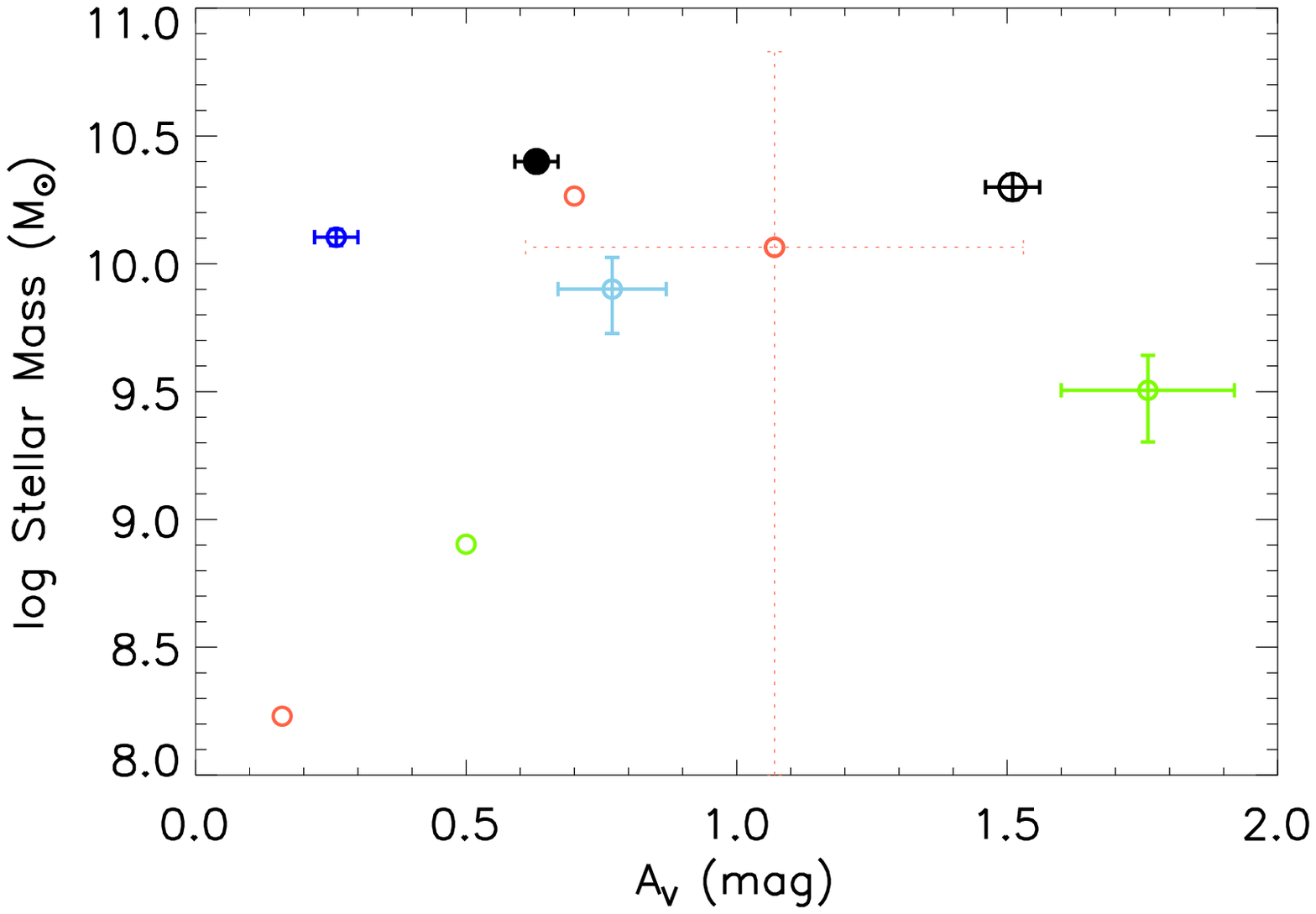,width=8.4cm}
\caption{Mean SED properties derived for individual LAEs in several publications at different redshifts; at $z \sim 3/3.7$ with a light blue/green point from Ono et al.~(2010a) and Finkelstein et al.~(2009a). Red points are at $z \sim 5.7$ from Lai et al. (2007), Pirzkal et al. (2007) and at $z \sim 4.9$ from Yuma et al.~(2010) and the dark blue point is the mean of a sample of $z \sim 0.3$ LAEs from Finkelstein et al. (2009b), with confirmed AGN excluded (Finkelstein et al.~2009c).
The black circles marks the mean in our sample of 40 best fit candidates; the filled circle is for the sub-sample with two populations and the open circle is for the single (young) population galaxies. }
\label{fig:properties2}
\end{center}
\end{figure}

In the interest of understanding any possible trends in the dust extinction with redshift we plot SED fitting results for the stellar mass and A$_V$ in Fig.~\ref{fig:properties2}.
In this plot, only fits done on individually detected objects are shown as it was found in sec.~\ref{sec:stackright} that the dust extinction cannot be well determined with stacked fits. Unfortunately, this leaves only a few points in the plot, and no real trends are visible.

Several difficult choices have to be made when preparing to fit SEDs of high redshift galaxies, besides the choice to stack or not, all of which may influence the final result (see also Gawiser et al.~2009). Firstly, the theoretical model must be chosen. We have here chosen the GALAXEV model, as this is most common in publications to date. Secondly, different groups fit different star formation histories (e.g. SSP, constant SFR, exponentially declining SFR) which will influence the results. As described in Sect.~\ref{sec:modelimprove}, nebular emission makes a strong contribution to the SED at very young ages, but this effect is typically not implemented in SED fitting codes. Some publications have included the narrow-band flux in the fitting, in an attempt to model the dust extinction (e.g. Finkelstein et al.~2008), whereas most groups tend to either ignore the data-point, or use it to subtract the Ly$\alpha$ flux from the broad-bands. Finally, one effect which may be the more difficult to circumvent, is that of disparate restframe wavelength coverage at different redshifts. At higher redshifts, the fits rely increasingly on the \emph{Spitzer} data points to constrain the restframe optical and near-infrared. The uncertainties there lie in the low spatial resolution, and that, at fainter fluxes, the images quickly over-crowd.

\section{Discussion}\label{sec:conclusions}
In this paper we have presented the results of fitting the SEDs of a set of Ly$\alpha$ emitters at $z=2.25$.
The results presented here represent the single largest sample of individually studied Ly$\alpha$ emitters to date. This is for the simple reason that galaxies at $z \sim 2$ are much more easily observable in the continuum than at higher redshifts. In Nilsson et al.~(2009a), we proposed that LAEs at $z\sim2$ are dustier and more massive than at higher redshift, and that they contain more AGN. These suggestions are largely reinforced with the results presented here. With the new \emph{Chandra} images, the number of detected AGN in the sample has increased to more than double what was found in Nilsson et al.~(2009a). The LAEs in this sample have a typical A$_V$ of $\sim 0.6$ and a typical stellar mass of a few times $10^{10}$~M$_{\odot}$. 

It is remarkable that almost all of the sample with a single young population have $A_V > 1.0$, whereas nearly all of the older objects have $A_V < 1.0$. This is a potentially interesting result, albeit potentially biased due to selection effects, e.g. that young low-dust galaxies will be faint in the bands redward of the Balmer break. Interestingly, no clear correlations were found between the Ly$\alpha$ properties of the galaxies, and their stellar masses or dust contents. The apparent, though with low significance, anti-correlation between Ly$\alpha$ flux and A$_V$, with fainter Ly$\alpha$ fluxes at larger A$_V$, is a classic result, indicating that the extinction in general affects the Ly$\alpha$ emission in a strong negative way. 

It may be argued that the results presented here are biased to the brightest end of the galaxy sample, in that only objects individually detected in the redder bands are fitted. For example, summing the average stellar mass found for the total stack, and comparing it to the sum of the individually fit objects (sec.~\ref{sec:indresults}), we find that those 30\% of the candidates contain 88\% of the total stellar mass among the galaxies. These 30\% are obviously among the brightest, although the secondary criteria of having small $\chi^2_r$ is also imposed. 
However, the same bias exists in all other similar campaigns to understand the properties of high redshift galaxies. For this reason we also chose to fit the stacked fluxes, and tested if the fitting results were consistent between stacks and individual fits. Of the three main parameters reported here, the stellar mass was the most robust and can be trusted in all situations. Ages were slightly lower when stacks were fitted, compared to individual results, but the parameter that is most incorrectly determined for a stack is the dust extinction. Due to this uncertainty, and the few results reported so far for individually fitted objects, it is difficult to draw any conclusions regarding dust content evolution with redshift.  
As for the stellar masses, the stacked average masses found at $z \sim 3$ are of the order $0.5 - 1.0 \times 10^9$~M$_{\odot}$, whereas the total stack here was best fitted with an average mass of $\sim 1 \times 10^{10}$~M$_{\odot}$, about a factor of ten larger. We have also shown here that our stacked flux limit in the red continuum bands is comparable to, or in many cases deeper than, the stacked limits in previous surveys, and hence that the increase in stellar mass with redshift is robust.

Further, we see no evidence for two populations of LAEs, with a red and a blue sub-sample. Instead, the results in Fig.~\ref{fig:results} follow a smooth transition from little to a large amount of dust, and from smaller to larger masses. If a flux cut was made in our sample at a higher flux, only the most massive objects would be individually detected. The stack of the remaining objects would be dominated by smaller mass objects and so if this stack is also fitted for, it would appear as though there was a dichotomy of LAE classes. The smooth transition of properties seen in the sample here rather indicates that Ly$\alpha$ emitters can be found in any type of galaxy at $z=2.25$.
Thus, Ly$\alpha$ emitters are not a homogeneous type of high redshift galaxy, certainly less so than i.e.~Lyman-Break galaxies (Nilsson et al. 2010). Selecting by Ly$\alpha$ emission will, and does, find a cross-section of Ly$\alpha$ emitting objects at each redshift and even though this technique is complementary to other high redshift galaxy selection techniques, the most valuable comparisons are probably with other galaxies at the same redshift. For example, Shapley et al.~(2005) found that LBGs at $z\sim2$ had stellar masses $\log$~M$_{\star} = 10.8$~M$_{\odot}$, A$_V = 0.86$ mag and ages around 1 Gyr, with 15\% having very young ages ($<100$~Myrs). These values are more consistent with the results found for LAEs here, than comparing LAE results at different redshifts. To conclude, Ly$\alpha$ emitters trace systematically different galaxies at different redshifts, rather than comprising a certain class of galaxies, and can thus be used as a tool to find interesting objects at high redshift.

\section{Conclusion}\label{sec:conclusion}
Based on the work presented here, studying the photometric properties of $z=2.25$ LAEs, and fitting the SEDs of the objects with spectral templates, the following conclusions can be made.
\begin{itemize}
\item[-] \emph{LAEs at $z=2.25$ are generally more massive and possibly more dusty than at higher redshift, and their properties are quite diverse}
\end{itemize}
Using a sophisticated MCMC fitting method, the stellar properties were derived for 40 well-fit objects, as well as three stacks. The LAEs were typically best fit with A$_V = 0.0 - 2.5$~mag and $\log $~M$_* = 8.5 - 11.0$~M$_{\odot}$. We have shown that when multiple populations of stars exist in the galaxies, the older populations are often outshone by the younger. Roughly $15$\% of the LAEs studied here are in such a strong star-bursting phase that any other older populations are invisible, whereas the remaining $85$\% have ``confirmed'' older populations of stars. In section~\ref{sec:discussion} we showed that LAEs at higher redshift ($z>3$) tend to have smaller masses.
\begin{itemize}
\item[-] \emph{Larger AGN fraction among LAEs}
\end{itemize}
An analysis of the \emph{Chandra} X-ray images available in COSMOS revealed 10 new AGN detections, increasing the AGN fraction in this LAE sample to $11$\%. 
\begin{itemize}
\item[-] \emph{Stacking of objects does not reveal the average of the properties of the individual objects}
\end{itemize}
SED fitting of high redshift galaxies is typically done on stacks of objects. We tested here the assumption that the stack reveals the average of the individual results. By stacking the individually best fit 40 objects in different ways, we showed that the fitted stellar properties of this stack were not the average of the properties of the individual objects. In the process of stacking, the light from multiple populations of stars is smeared out, and particularly the A$_V$ determination becomes highly uncertain. 
\begin{itemize}
\item[-] \emph{Strong correlations between ages, stellar masses and redshift were seen}
\end{itemize}
A correlation between the age and stellar masses with $99.94$\% probability was detected when comparing SED fitting results for LAEs at different redshifts. Further, a $99.92$\% probability correlation was found between age and redshift. This agrees well with a scenario where lower redshift LAEs are older galaxies experiencing a second burst of star formation.

\begin{acknowledgements}
The Dark Cosmology Centre is funded by the DNRF. G{\"O} is a Royal Swedish Academy of Sciences research fellow, supported through a grant from the Knut and Alice Wallenberg foundation, and also acknowledges support from the Swedish research council and the Swedish national space board.
\end{acknowledgements}

\end{document}